\documentclass{aa}  

\usepackage{graphicx}
\usepackage{txfonts}
\usepackage{natbib}
\usepackage{placeins}
\usepackage{hyperref}
\usepackage{graphicx}
\usepackage{amssymb}
\usepackage{amsmath}

\let\tablenum\relax
\usepackage{siunitx} 
\newcommand{\um}{\SI{}{\micro\meter}}

\newcommand\jwst{\textit{JWST}}
\newcommand\hst{\textit{HST}}

\newcommand{\re}{R$_{\rm e}$}
\newcommand{\zr}[2]{$z$\,$\approx$\,{#1}\,--\,{#2}}

\def\arcmin{\ifmmode {^{\prime}}\else $^{\prime}$\fi}
\def\arcsec{\ifmmode {^{\prime\prime}}\else $^{\prime\prime}$\fi}
\newcommand\qth{$^{\rm th}$}
\newcommand{\av}{$A_{\rm V}$}
\newcommand{\orcid}[1]{\includegraphics[scale=0.06]{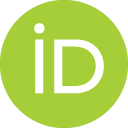} \href{https://orcid.org/#1}{#1}}

\begin{document} 

   \title{Sub-Millimetre Galaxies with Webb:}
   \subtitle{Near-Infrared Counterparts and Multi-wavelength Morphology}

   \author{Steven Gillman,\inst{1,2}
   \thanks{\href{orcids}{ORCIDs} listed on final page}
          Bitten Gullberg, \inst{1,2}
          Gabe Brammer,  \inst{1,3}
          Aswin P. Vijayan, \inst{1,2}
          Minju Lee,  \inst{1,2} 
          David Blánquez,  \inst{1,2}
          Malte Brinch, \inst{1,2}
          Thomas R. Greve,  \inst{1,2,4}
          Iris Jermann,      \inst{1,2} 
          Shuowen Jin,  \inst{1,2}
          Vasily Kokorev, \inst{5}
          Lijie Liu,  \inst{1,2}
          Georgios Magdis,  \inst{1,2,3}
          Francesca Rizzo,  \inst{1,3}
          \and 
          Francesco Valentino  \inst{1,3,6}
          }
   \institute{Cosmic Dawn Center (DAWN), Denmark\\
    \email{srigi@space.dtu.dk}
    \and
        DTU-Space, Elektrovej, Building 328 , 2800, Kgs. Lyngby, Denmark
    \and
        Niels Bohr Institute, University of Copenhagen, Jagtvej 128, DK-2200 Copenhagen N, Denmark
    \and
    Dept.~of Physics and Astronomy, University College London, Gower Street, London WC1E 6BT, United Kingdom
    \and 
    Kapteyn Astronomical Institute, University of Groningen, P.O. Box 800, 9700AV Groningen, The Netherlands
    \and 
    European Southern Observatory, Karl-Schwarzschild-Str. 2, D-85748 Garching bei Munchen, Germany
        }

   \date{Received \today; accepted \today}

 
  \abstract
{We utilise the unprecedented depth and resolution of recent early-release science (ERS) \jwst{} observations to define the near-infrared counterparts of sub-millimetre selected galaxies (SMGs). We identify 45 SCUBA-2 SMG positions within The Cosmic Evolution Early Release Science Survey (CEERS) \jwst/NIRCam fields. Through an analysis of multi-wavelength $p$-values, NIRCam colours and predicted SCUBA-2 fluxes, we define 43 \jwst{}/NIRCam counterparts to the SCUBA-2 SMGs, finding a 63 per cent agreement with those identified in prior \hst{} studies. Using \texttt{EaZy-py} we fit the available \hst{} and \jwst{} observations to quantify the photometric redshifts of the NIRCam-SMGs, establishing a broad range of redshift from $z$\,$\approx$\,0.2\,--\,5.4 with a median of $z$\,$\approx$\,2.29, in agreement with other studies of SMGs. We identify significant variations in the morphology of the NIRCam-SMGs from isolated discs and spheroidal galaxies  to irregular interacting systems. We analyse their rest-frame optical and near-infrared morphological properties (e.g. effective radius (\re), S\'ersic index ($n$), Concentration ($C$), Asymmetry ($A$), Clumpiness ($S$) as well as the Gini and  M$_{20}$ parameters), finding, on average, late-type disc-like morphologies with large scatter into the intermediate and merger regions of the non-parametric parameter space. For the non-merging galaxies, we find a median rest-frame optical size and S\'ersic index (and $1\sigma$ scatter) of R$_{\rm e}$\,=\,3.10\,$\pm$\,1.67\,kpc and $n$\,=\,0.96\,$\pm$\,0.66. Whilst in the rest-frame near-infrared we establish more compact, higher S\'ersic index morphologies (R$_{\rm e}$\,=\,1.64\,$\pm$\,0.97, $n$\,=\,1.85\,$\pm$\,0.63). We further establish that both the rest-frame optical and near-infrared effective radii correlate negatively (at a 2$\sigma$ level) with redshift whilst the  S\'ersic index remains constant with cosmic time. Our results are consistent with the picture of inside-out galaxy evolution, with more centrally concentrated older stellar populations, and more extended, younger star-forming regions whose stellar emission is heavily attenuated in the central regions.
}
             
\titlerunning{Sub-Millimetre Galaxies with Webb}
\authorrunning{S. Gillman at al.}
\maketitle

%

\section{Introduction} \label{sec:intro}

Sub-millimetre galaxies (SMGs), defined by their excess ($\gtrsim$\,1\,--\,2\,mJy) sub-millimetre (450, 850\um) flux, were first identified in ground-based wide-field single-dish observations from telescopes such as SCUBA \citep[e.g.][]{Smail1997,Eales1999}\footnote{We note SMGs are often also classified as IR-selected dusty star-forming galaxies (DSFG), (see \citet{Casey2014} for review), but here we focus on the original single-dish selected definition of SMGs.}. In the last few decades, continued analysis has shown these sub-millimetre bright systems constitute a unique, massive \citep[$\sim$\,10$^{11}$M$_{\odot}$;][]{Wardlow2011,Simpson2014} galaxy population that is heavily dust-obscured and far-infrared (FIR) luminous, commonly exhibiting intense star formation rates \citep[$\sim$10$^{2}$\,--\,10$^{3}$\,M$_{\odot}$yr$^{-1}$;][]{Blain2002,Chapman2005,Swinbank2014}.
 
Reproducing these extremely dusty, massive systems in simulations has been a long-standing problem \citep[e.g.][]{Baugh2005,Dave2010,Mcalpine2019}. Often simulations have to be tuned to reproduce the sub-millimetre number counts and redshift distribution, invoking novel modelling assumptions (e.g. top-heavy initial mass function \citealt{Cowley2019}). More recent simulations with complex dust models or sub-millimetre flux scaling relations \citep[e.g.][]{Lovell2021,Hayward2021} have had more success and established that the sub-millimetre bright phases are a natural consequence of massive galaxy evolution in a $\Lambda$CDM Universe \citep[e.g.][]{Long2022,Cochrane2022,Lower2022}. Despite these recent successes, constraining the multi-wavelength properties of sub-millimetre bright galaxies across cosmic time still poses many observational challenges due to their inherent faintness at optical to infrared wavelengths \citep[e.g.][]{Dey1999,Weiss2009,Hodge2012,Smail2021}.

To ascertain the cosmic evolution of SMGs, their full multi-wavelength (UV to far-infrared) properties need to be quantified. In the far-infrared, wide-field single-dish observations, although often incredibly deep (1$\sigma$ depth  of 2\,mJy\,beam$^{-1}$; \citealt{Smail1997}) which is crucial for identifying SMGs, suffer from large beam sizes and poor resolution. For instance the Sub-millimetre Common-User Bolometer Array 2 \citep[SCUBA-2;][]{Dempsey2013} at 850\um\ has a 14.5\arcsec FWHM \citep{Holland2013}. Thus identifying the individual multi-wavelength counterparts to the sub-millimetre selected galaxies is very challenging \citep[e.g.][]{Ivison2007,Biggs2011,Chen2016,An2018}. This is compounded at optical\,--\,near-infrared wavelengths where SMGs are inherently faint due to their extremely dusty nature, and telescopes such as the \textit{Hubble Space Telescope} (\hst) provide sub-arcsecond resolution revealing a multitude of potential counterparts \citep[e.g.][]{Zavala2018,Ling2022,Shim2022}. 

In recent years, high-resolution continuum observations from the Atacama Large Millimetre/sub-millimetre Array (ALMA) have followed up single-dish surveys, identifying individual galaxy counterparts at sub-millimetre wavelengths (see \citealt{Hodge2020} for a review of ALMA SMG studies). A large fraction of these observations identify that SMGs are commonly residing in over-dense regions \citep[e.g.][]{Simpson2014,Hodge2016, Miller2018, Oteo2018}, with 10\,--\,80$\%$ of bright single-dish sub-millimetre sources found to comprise of two or more individual SMGs when observed at sub-mJy rms depths \citep[e.g.][]{Karim2013,Simpson2014,Stach2018,An2019}. A single bright sub-millimetre  source, however, can comprise of multiple SMGs over a broad range of redshifts. 
Both the negative \textit{K}-correction, which effectively fixes the luminosity as a function of redshift \citep[e.g.][]{Hill2018}, or alternatively strong gravitational lensing \citep[e.g.][]{Weiss2013,Diaz2017,Harrington2021} can contribute to this `line of sight' blending. 

Studies such as \citealt{Hodge2016,Gullberg2019,Cochrane2021} which utilised sub-arcsecond resolution ($\leq$0\farcs{2}) observations with ALMA, have revealed compact ($\sim$few kpc) dusty galaxies, often with observed-frame far-infrared disc-like morphologies (i.e $n$\,$\approx$\,1) and kinematics \citep[e.g.][]{Lelli2021,Rizzo2021}. In contrast, rest-frame ultra-violet \hst{} studies of \zr{1}{3} SMGs identify extended (R$_{\rm dust}$/R$_{\rm stellar}$\,$\approx$\,0.6) irregular morphologies with multiple components \citep[e.g.][]{Swinbank2010,Chen2015,Gomez2018,Zavala2018,Lang2019,Ling2022}. 

Detecting and resolving SMGs rest-frame near-infrared emission, which reflects the bulk of the stellar population, has however, previously been incredibly  challenging. Observations with the Infrared Array Camera (IRAC; \citealt{Fazio2004}) on the \textit{Spitzer} space telescope, whilst providing the required near-infrared wavelength coverage from 3.6\,--\,8\um, lacks the spatial (sub-arcsecond) resolution required to constrain the underlying morphologies of the stellar emission in SMGs \citep[e.g.][]{Krick2021}. With the advent of recent high-resolution observations from \jwst{}, the near-infrared counterparts of sub-millimetre bright galaxies can be identified and their subsequent properties (e.g. stellar morphology) quantified. With \jwst{}  observations we can robustly constrain the total mass budget of this unique galaxy population and constrain their evolution across cosmic time.

To this end, in this paper, we present an analysis of the infrared counterparts to SCUBA-2 450 and 850\um\ selected SMGs detected in public \jwst{} observations  from the early-release science program; The Cosmic Evolution Early Release Science Survey (CEERS; Program ID:1345, PI: Finkelstein; \citealt{Bagley2022})\footnote{\url{https://ceers.github.io}}. In Section \ref{sec:sample}, we define the sample of SCUBA-2 SMGs that have been observed with \jwst{} and compile the public multi-wavelength data. Through an analysis of the near-infrared NIRCam colours, multi-wavelength (850\um, 24\um, 4.44\um) $p$-values and predicted spectral energy distribution (SED) fluxes, we first identify the NIRCam counterparts to the SMGs in Section \ref{Sec:counterparts} before analysing their  rest-frame multi-wavelength morphology in Section \ref{sec:morph} and summarising our analysis in Section \ref{Sec:conc}. Throughout the paper, we assume a $\Lambda$CDM cosmology with $\Omega_{\rm m} = 0.3$, $\Omega_{\Lambda} = 0.7$, and $H_0 = 70\,\mathrm{km\,s^{-1}\,Mpc^{-1}}$. All quoted magnitudes are on the AB system and stellar masses are calculated assuming a Chabrier initial mass function (IMF) \citep{Chabrier2003}. 

\section{Sample Selection}\label{sec:sample}
To identify the near-infrared counterparts to the SMGs observed with \jwst{}, we first compile a list of previously characterised SMGs in the literature, including samples from \citealt{Zavala2017,Geach2017}. We then cross-match this sample with the JWST/NIRCam observations from the Cycle 1 ERS program CEERS, resulting in 45 SCUBA-2 pointings falling within the field of view of the NIRCam observations. This sample of 45 is comprised of both 850\um\ and 450\um\ SCUBA-2 pointings from the \citet{Zavala2017} sample, as we describe in the next section and present in Table \ref{Table:Targets}.

\begin{table*}
\centering
\caption{Table summarising the SCUBA-2 SMGs identified in CEERS Epoch 1 NIRCam Imaging. The 850, 450 IDs, SCUBA-2 RA and Dec. and 850, 450 S/N and deboosted fluxes (columns 1\,--\,8) are from \citet{Zavala2017}. Column 9 indicates if an optical ({\em HST}/F160W) counterpart is identified in \citet{Zavala2018} and column 10 indicates if a 1.4\,GHz radio counterpart is reported in \citet{Ivison2007}.}
\begin{tabular}{cccccccccc}
\hline
 850 ID & 450 ID & RA & Dec.  & Deboosted $S_{\rm 850}$ & S/N$_{\rm 850}$ &  Deboosted $S_{\rm 450}$ & S/N$_{\rm 450}$ & {\em HST} & Radio \\
         &        &  SCUBA-2 & SCUBA-2 & (mJy\,beam$^{-1}$) &  &   (mJy\,beam$^{-1}$)         &                 &   1.6\um\     &  1.4GHz \\
\hline
850.003 & 450.05 & 214.917 & 52.891 & 5.24\,$\pm$\,0.31 & 23.2 & 13.3\,$\pm$\,2.0 & 9.9 & \checkmark  & \checkmark  \\
850.004 & 450.25 & 214.947 & 52.910 & 3.87\,$\pm$\,0.31 & 17.0 & 6.0\,$\pm$\,2.2 & 4.7 & \checkmark  &  \\
850.017 & 450.11 & 214.900 & 52.852 & 2.70\,$\pm$\,0.37 & 10.1 & 9.7\,$\pm$\,2.4 & 6.5 & \checkmark  &  \\
850.019 & 450.41 & 214.972 & 52.958 & 3.43\,$\pm$\,0.51 & 9.6 & 7.3\,$\pm$\,3.5 & 3.9 & \checkmark  &  \\
850.025 & 450.29 & 214.852 & 52.866 & 2.68\,$\pm$\,0.42 & 9.0 & 6.6\,$\pm$\,2.9 & 4.5 & \checkmark  &  \\
850.026 & 450.31 & 214.890 & 52.894 & 2.40\,$\pm$\,0.38 & 8.8 & 5.9\,$\pm$\,2.6 & 4.3 &  &  \\
850.028 & 450.69 & 214.878 & 52.853 & 2.45\,$\pm$\,0.42 & 8.2 & 3.9\,$\pm$\,2.5 & 3.2 & \checkmark  &  \\
850.030 & 450.10 & 214.879 & 52.877 & 2.14\,$\pm$\,0.38 & 8.0 & 11.1\,$\pm$\,2.4 & 7.2 & \checkmark  & \checkmark  \\
850.031 & 450.49 & 214.823 & 52.873 & 2.87\,$\pm$\,0.36  & 8.0  & 6.0\,$\pm$\,3.3 & 3.8 &  &  \\
850.034 & 450.66 & 214.916 & 52.952 & 2.14\,$\pm$\,0.40 & 7.5 & 4.3\,$\pm$\,2.8 & 3.3 &  &  \\
850.038 & 450.27 & 214.865 & 52.899 & 2.08\,$\pm$\,0.41 & 7.2 & 7.1\,$\pm$\,2.9 & 4.6 & \checkmark  &  \\
850.043 & 450.73 & 214.950 & 52.937 & 1.68\,$\pm$\,0.35 & 6.9 & 3.6\,$\pm$\,2.4 & 3.1 & \checkmark  &  \\
850.059 & 450.24 & 214.857 & 52.849 & 1.62\,$\pm$\,0.51 & 4.9 & 8.3\,$\pm$\,3.2 & 4.8 & \checkmark  &  \\
850.065 & 450.18 & 214.877 & 52.867 & 1.35\,$\pm$\,0.38 & 5.1 & 8.4\,$\pm$\,2.5 & 5.5 & \checkmark  & \checkmark  \\
850.070 & 450.34 & 214.970  & 52.930  & 1.15\,$\pm$\,0.34 & 4.9 & -   & -  & \checkmark  &  \\
850.027 & - & 214.910 & 52.937 & 2.25\,$\pm$\,0.36 & 8.7 & -   & -  &  &  \\
850.035 & - & 214.901 & 52.943 & 2.07\,$\pm$\,0.39 & 7.5 & -   & -  &  &  \\
850.036 & - & 214.905 & 52.922 & 1.89\,$\pm$\,0.36 & 7.3 & -   & -  &  &  \\
850.047 & - & 214.859 & 52.861 & 1.91\,$\pm$\,0.42 & 6.5 & -   & -  & \checkmark  &  \\
850.054 & - & 214.874 & 52.843 & 1.95\,$\pm$\,0.51 & 5.7 & -   & -  & \checkmark  &  \\
850.055 & - & 214.880 & 52.912 & 1.56\,$\pm$\,0.39 & 5.6 & -   & -  &  &  \\
850.058 & - & 214.848 & 52.852 & 1.89\,$\pm$\,0.51 & 5.6 & -   & -  &  &  \\
850.063 & - & 214.867 & 52.883 & 1.40\,$\pm$\,0.39 & 5.2 & -   & -  &  &  \\
850.064 & - & 214.947 & 52.924 & 1.19\,$\pm$\,0.33 & 5.2 & -   & -  &  &  \\
850.067 & - & 214.831 & 52.893 & 1.55\,$\pm$\,0.46 & 5.1 & -   & -  & \checkmark  &  \\
850.068 & - & 214.893 & 52.933 & 1.30\,$\pm$\,0.38 & 5.0 & -   & -  &  &  \\
850.075 & - & 214.835 & 52.869 & 1.26\,$\pm$\,0.46 & 4.3 & -   & -  & \checkmark  &  \\
850.076 & - & 214.917 & 52.956 & 0.83\,$\pm$\,0.50 & 3.3 & -   & -  &  &  \\
850.080 & - & 214.925 & 52.934 & 0.91\,$\pm$\,0.36 & 4.0 & -   & -  & \checkmark  &  \\
850.086 & - & 214.945 & 52.929 & 0.60\,$\pm$\,0.36 & 3.1 & -   & -  &  &  \\
850.091 & - & 214.900 & 52.838 & 0.94\,$\pm$\,0.71 & 3.6 & -   & -  &  &  \\
850.093 & - & 214.930 & 52.920 & 0.72\,$\pm$\,0.38 & 3.5 & -   & -  &  &  \\
850.098 & - & 214.923 & 52.928 & 0.69\,$\pm$\,0.39 & 3.4 & -   & -  &  &  \\
850.099 & - & 214.944 & 52.946 & 0.73\,$\pm$\,0.41 & 3.3 & -   & -  &  & \checkmark  \\
850.101 & - & 214.838 & 52.859 & 0.83\,$\pm$\,0.72 & 3.3 & -   & -  &  &  \\
850.108 & - & 214.866 & 52.871 & 0.68\,$\pm$\,0.43 & 3.1 & -   & -  &  &  \\
- & 450.35 & 214.904 & 52.863 & -   & -  & 5.2\,$\pm$\,2.4 & 4.2 &  &  \\
- & 450.48 & 214.910 & 52.927 & -   & -  & 4.9\,$\pm$\,2.8 & 3.8 & \checkmark  &  \\
- & 450.68 & 214.957 & 52.947 & -   & -  & 4.0\,$\pm$\,2.6 & 3.2 &  &  \\
- & 450.70 & 214.906 & 52.933 & -   & -  & 3.9\,$\pm$\,2.5 & 3.2 &  &  \\
- & 450.42 & 214.905 & 52.851 & -   & -  & 4.6\,$\pm$\,2.5 & 3.8 & \checkmark  & \checkmark  \\
- & 450.77 & 214.844 & 52.894 & -   & -  & 3.8\,$\pm$\,2.5 & 3.1 &  &  \\
- & 450.81 & 214.934 & 52.906 & -   & -  & 3.1\,$\pm$\,1.9 & 3.1 &  &  \\
- & 450.88 & 214.875 & 52.888 & -   & -  & 3.4\,$\pm$\,2.3 & 3.0 &  &  \\
- & 450.89 & 214.916 & 52.852 & -   & -  & 3.1\,$\pm$\,1.9 & 3.0 &  &  \\
\hline
\end{tabular}
\label{Table:Targets}
\end{table*}

\subsection{SCUBA-2 Observations}

We utilise the sample of SMGs presented in \citet{Zavala2017} which were observed with SCUBA-2 camera on the James Clark Maxwell Telescope (JCMT) between 2012 and 2015 as part of the SCUBA-2 Cosmic Legacy Survey \citep[S2CLS;][]{Geach2017} in the extragalactic deep field Extended Growth Strip (EGS; \citealt{Davis2007}). The SMG sample defined in \citet{Zavala2017} derive from the `deep tier' observations of the survey, whilst the `wider tier' observations are presented in \citet{Geach2017}. For details of the observations, data reduction and SMG identification we refer the reader to \citet{Zavala2017} and references therein.

In short, observations were carried out at 450 and 850\um\ in the EGS field with 8 and 14.5\arcsec{} FWHM beams respectively. The 450 (850) positions, S/N and fluxes of the SMGs are reported in Appendix A of \citet{Zavala2017} and summarised in Table \ref{Table:Targets}. We use these target positions and 450 (850) beam sizes as positions and search radii, respectively, within \jwst/NIRCam observations to identify near-infrared counterparts to the SCUBA-2 SMGs. If the SCUBA-2 SMG is detected in both 450 and 850\um\ observations, we use the 450 position and beam due to its smaller FWHM. The 850\um\ selected SMGs have a deboosted median (and 16\qth\,--\,84\qth{} quartile) flux of S$_{\rm 850}$\,=\,1.62$^{+2.43}_{-0.83}$\,mJy\,beam$^{-1}$ whilst the 450\um\ selected SMGs have a median (and 16\qth\,--\,84\qth{} quartile) deboosted flux of S$_{\rm 450}$\,=\,5.2$^{+8.35}_{-3.70}$\,mJy\,beam$^{-1}$.

\subsection{\hst{} and \jwst/NIRCam Observations}\label{Sec:JWST}
 
To identify the near-infrared counterparts of the 45 SMGs outlined above we analyse the CEERS Epoch 1 \jwst/NIRCam imaging. The NIRCam imaging covers 34.7 sq.~arcmin of the EGS field with observations taken in both short wavelength (F115W, F200W) and long wavelength (F277W, F356W, F444W) filters. The level-2 data products were downloaded from STScI portal\footnote{\url{https://mast.stsci.edu}} and processed using \texttt{Grizli} pipeline \citep{Brammer2021,Brammer2022} with routines to derive accurate NIRCam photometric zeropoints and correct for cosmic rays and stray light \citep[e.g.][]{Boyer2022,Nardiello2022,Bradley2022}. Further incorporating available optical and near-infrared \hst{} observations from the Complete \textit{Hubble} Archive for Galaxy Evolution \citep[CHArGE][]{Kokorev2022}, the final images were aligned to Gaia DR3 \citep{Gaia2021} and drizzled \citep{Fruchter2002} to a resolution of $0.04\arcsec$ per pixel. 

\begin{figure*}[ht!]
    \centering
    \includegraphics[width=\linewidth,trim={0 0cm 0 0cm},clip]{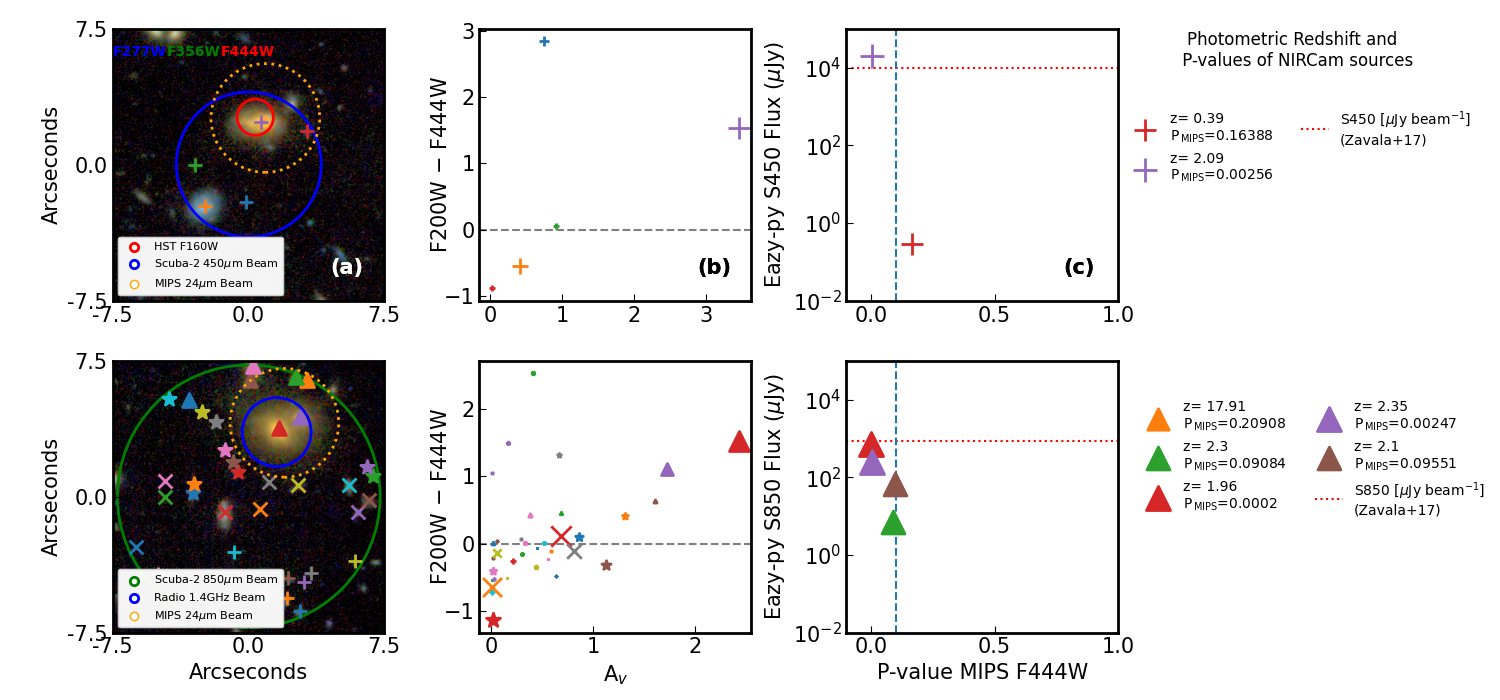}
    \caption{Two examples of the multi-wavelength data used to identify the NIRCam-SMGs for 450\um{}  (\textbf{upper}) and 850\um{} (\textbf{lower}) detected SCUBA-2 SMGs. We show a F277W, F356W, F444W three colour NIRCam image centred on the SCUBA-2 position \textbf{(a)}. In panel \textbf{b} we correlate the NIRCam F200W-F444W colour with the \texttt{EaZy-py} \av\ values for the NIRCam sources, where the marker size reflects the MIPS to F444W $p$-value of the NIRCam source (lower $p$-value, larger marker) and the grey-dashed line indicates a colour of zero. Finally in panel \textbf{c} the \texttt{Eazy-py} predicted fluxes as a function of MIPS to F444W $p$-values are shown. We also indicate the classical $p$-value\,$<$\,0.1 threshold (vertical blue-dashed line) and observed 450\um\ (850\um) deboosted SCUBA-2 fluxes (red-dashed horizontal lines). NIRCam SMGs are identified, following the decision tree (Appendix \ref{App:A}), by having high \av, red colour, low $p$-values and \texttt{EaZy-py} predicted fluxes comparable to the SCUBA-2 fluxes in \citet{Zavala2017}.}
    \label{Fig:counterparts}
\end{figure*}

Sources were then identified using \texttt{SEP}\footnote{\url{https://github.com/kbarbary/sep}} \citep{Barbary2016}, a python-wrapped \texttt{Source Extractor} \citep{Bertin1996} run on a noise-weighted combined long-wavelength channel NIRCam image (F277W+F356W+F444W). Aperture photometry was performed on each source using 0\farcs{5} circular apertures corrected to `total' values within an elliptical Kron aperture \citep{Kron1980}\footnote{The photometry catalogue and mosaics are available online \href{https://doi.org/10.17894/ucph.e3d897af-233a-4f01-a893-7b0fad1f66c2}{here.}}. For each source identified in the CEERS Epoch 1 observations, we run the \texttt{EaZy-py}\footnote{\url{https://github.com/gbrammer/eazy-py}} \citep{Brammer2008} SED fitting code on the available \hst{} and \jwst{} photometry to derive photometric redshifts. We use thirteen templates from the Flexible Stellar Populations Synthesis code \citep[FSPS;][]{Conroy2010} described in \citet{Kokorev2022} linearly combined to allow for maximum flexibility. We note however given the uncertainties associated with SED fitting \citep[e.g.][]{Pac2022} we only focus on the photometric redshift, dust attenuation and 450 and 850\um\ predicted fluxes for our analysis.  Further defining the full SED of the NIRCam-SMGs will be presented in future papers. We now use the positions, fluxes, and photometric redshift information for all the sources in CEERS NIRCam observations, in addition to the ancillary multi-wavelength data described in the next section, to identify the SMG counterparts.

\subsection{Ancillary Data}

In addition to the \jwst/NIRCam and SCUBA-2 observations detailed in prior sections, we take advantage of other publicly available multi-wavelength observations of the EGS field to aid the identification of the near-infrared counterparts of the SMGs.

For each of the SCUBA-2 positions outlined in Table \ref{Table:Targets}, we compile the available Radio 1.4\,GHz data products from the All Wavelength Extended Groth Strip International Survey (AEGIS20; \citealt{Ivison2007}) and MIPS 24\um\ data products (images and source catalogue) from the  Far-Infrared Deep Extragalactic Legacy Survey (FIDEL; PI: Dickinson, \citealt{Dickinson2007}). We identify five sources with 1.4\,GHz counterparts to the sub-millimetre detection, within the SCUBA-2 (450/850) beam as listed in Table \ref{Table:Targets}. All the SCUBA-2 SMG pointings are identified to have MIPS 24\um\ imaging with the majority (40/45) having 24\um\ sources identified in the SCUBA-2 beam, as derived by the FIDEL consortium\footnote{\url{http://benjaminmagnelli.weebly.com/downloads.html}}. Using this multi-wavelength dataset for each of the SCUBA-2 positions we proceed to first identify the near-infrared counterpart of the SMGs and then measure their multi-wavelength morphological properties.

\section{Near-Infrared Counterparts}\label{Sec:counterparts}

Defining the near-infrared counterparts of the SMGs requires us to analyse all the available multi-wavelength data for each SCUBA-2 SMG. Firstly, for the SCUBA-2 SMG positions detailed in Table \ref{Table:Targets}, we make $15\arcsec\times 15\arcsec$ NIRCam F444W cutouts centred on the 850\um\ or 450\um\ SCUBA-2 source position. As shown in Figure \ref{Fig:counterparts}, we then overlay all the potential NIRCam sources which fall within the beam. Using a combination of $p$-values, near-infrared colours, \texttt{EaZy-py} predicted SCUBA-2 fluxes and dust attenuation \av{}, as described in the following sections, we define a decision tree to identify the near-infrared counterparts of the SMGs (see Appendix \ref{App:A}).

\subsection{NIRCam Colours and Dust Attenuation}
It is well known that SMGs will exhibit very red optical-to-infrared colours due to their extremely dusty nature \citep[e.g.][]{Dey1999,Weiss2009,Walter2012}. Thus, to aid our identification of the SMGs in the \jwst/NIRCam observations, we calculate the F200W\,--\,F444W colour for each potential counterpart within the SCUBA-2 beam. We correlate this NIRCam colour with the dust attenuation (\av) from \texttt{EaZy-py} SED fitting as detailed in Section \ref{Sec:JWST}. Given the degeneracy between infrared colour and dust attenuation, we expect the SMGs' NIRCam counterparts to occupy the top-right of this plane with `red' NIRCam colour and high \av{} value \citep[e.g.][]{Martis2022}. We show an example of the correlation between F200W\,--\,F444W colour and \av{} in Figure \ref{Fig:counterparts}. 

\begin{figure*}[h!]
    \centering
    \includegraphics[width=\linewidth]{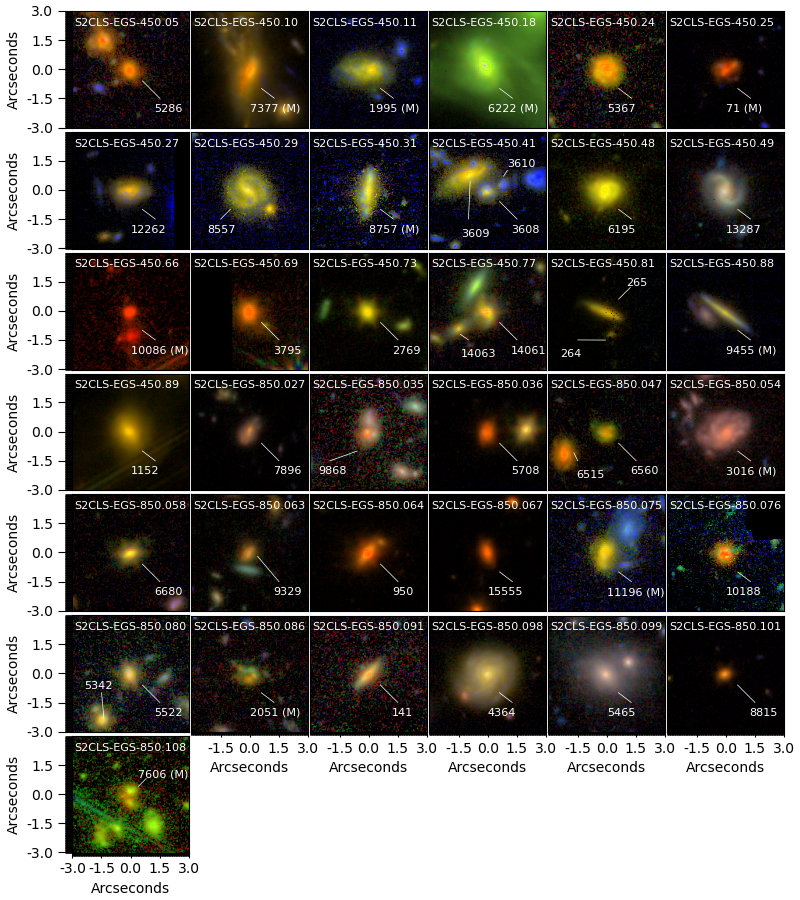}
    \caption{We show $6\arcsec \times 6\arcsec$ three-band (F115W/F200W, F277W/F356W, F444W) colour images for each SCUBA-2 SMGs with the NIRCam-SMGs labelled. In each image we indicate the ID of the NIRCam counterpart(s) with 1, 4, and 32 SCUBA-2 SMGs having 3, 2 and 1 NIRCam counterparts respectively. It is clear that the sample of NIRCam-SMGs contains a broad range of morphologies from isolated disc and spheroidal galaxies to disturbed interacting systems. We indicate the 11 galaxies classified non-parametrically (and visually) as Mergers  from Figure~\ref{Fig:non-para} with an (M).}
     \label{Fig:col_img}
\end{figure*}

\begin{figure}[ht]
    \centering
    \includegraphics[width=0.9\linewidth]{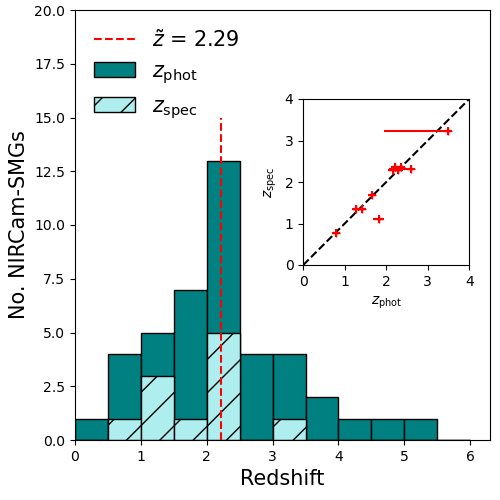}
    \caption{The photometric redshift distribution from \texttt{EaZy-py} for the NIRCam counterparts, with a median redshift of $\tilde{z}$\,=\,2.29 and range from \zr{0.21}{5.4}. The inset shows the spectroscopic redshift, where available, compared to the photometric redshift (and 16\qth\,--\,84\qth percentiles from the p($z$)). We derive a median ratio (and 16\qth\,--\,84\qth percentiles) of $z_{\rm phot}$/$z_{\rm spec}$\,=\,1.0$^{+1.10}_{-0.94}$.}
    \label{Fig:EazY_prop}
\end{figure}

\subsection{Multi-wavelength $p$-values and Predicted Fluxes}
Classically the SMG-counterpart selection is achieved through $p$-value calculations. This method is typically used to identify radio or far-infrared counterparts. In addition to the well-known FIR-radio correlation \citep[e.g.][]{Barger2014}, the FIR source density is often low and the FWHM of the observations is $\sim$\,few arcseconds, thus reducing the probability of source mis-identification.  First introduced in \citet{Downes1986}, the $p$-value corresponds to the probability the counterpart identified is a `by-chance' association. The $p$-value is a function of the magnitude of the potential counterpart and its proximity to the beam centre, whereby brighter and closer counterparts will have lower $p$-values. Typically counterparts with $p(m,r)<0.1$ are used as the selection criteria \citep[e.g.][]{Ivison2007,Biggs2011,Zavala2018}.

For our analysis, we calculate the $p$-value to go from the SCUBA-2 450\um{} (850\um) beam to MIPS 24\um\ imaging and then MIPS 24\um\ to F444W NIRCam imaging. For the five sources without MIPS coverage we calculate the $p$-value to go from SCUBA-2 450\um{} (850\um) to F444W NIRCam directly. In addition to the $p$-values of the NIRCam sources, we can also use their predicted 450\um{} (850\um) SCUBA-2 fluxes from the \texttt{EaZy-py} code (Section \ref{Sec:JWST}), as further information in the counterpart identification\footnote{We note however, these can be uncertain as \texttt{EaZy-py} assumes an energy balance for FIR. Discussion of this will be presented in future papers with more sophisticated SED fitting codes.}. For each SMG position, we plot the \texttt{EaZy-py} predicted SCUBA-2 450\um{} (850\um) fluxes as a function of the MIPS 24\um\ to F444W $p$-values, an example of which is shown in Figure \ref{Fig:counterparts}. We also overlay the observed SCUBA-2 450\um{} (850\um) fluxes from \citet{Zavala2017}. We expect the NIRCam counterpart to the SCUBA-2 SMG to exhibit low $p$-values with SCUBA-2 fluxes comparable to those reported in \citet{Zavala2017}. However, we do not use $p$-values as the only criteria for the NIRCam-SMG selection given the sub-arcsecond resolution of the \jwst{} observations.

\subsection{Near-Infrared Counterparts}

To identify the near-infrared counterpart(s) in the NIRCam F444W imaging, we construct a decision tree as shown in Appendix \ref{App:A}. The decision tree is designed to consider all of the available information for the SCUBA-2 positions, including the $p$-values, NIRCam colours, \texttt{EaZy-py} outputs (fluxes, \av) and MIPS coverage to define the most likely SMG counterpart. To make this process more robust, our team have made independent selections of the 45 SMG counterpart(s) using the information shown in Figure \ref{Fig:counterparts} and following the decision tree. We remove eight SCUBA-2 SMGs from the sample, due to poor multi-wavelength data resulting in inconclusive counterpart selection. From the remaining 37 SCUBA-2 SMGs, we identify 43 NIRCam counterparts with 1, 4 and, 32 SCUBA-2 SMGs having 3, 2 and, 1 NIRCam counterparts respectively. Of the 43 NIRCam counterparts, 23 have 100 per cent agreement in all independent selections. We label these quality 1 (Q1). Counterparts with $\geq$\,80 percent agreement (12/43) are labelled quality 2 (Q2) and $<$\,80 percent as quality 3 (Q3) (8/43). In the following analysis, we focus on Q1 and Q2 counterparts, but note that the inclusion of Q3 counterparts does not alter our conclusions. In Figure \ref{Fig:col_img} we show $6\arcsec\times 6\arcsec$ NIRCam colour images at the positions of the SCUBA-2 SMGs with the NIRCam-SMGs labelled by source ID, which in a number of cases contain multiple NIRCam sources. 

For just under half (19/43) of the SCUBA-2 SMGs in our sample, the optical counterparts have previously been identified in \hst{} F160W observations using similar analysis techniques (see \citealt{Zavala2018}). To validate our counterpart selection we compare to those identified in \citet{Zavala2018}, finding agreement with 63 per cent of previously identified sources. We note that given the increased depth and sensitivity of the \jwst/NIRCam observations compared to {\em HST}/WFC3 used in the \citet{Zavala2018} selection, as well as the difference in observed wavelength (4.44\um\ compared to 1.6\um), our selection is more sensitive to optically faint, redder objects that are not identified in \citet{Zavala2018}. Furthermore, studies indicate that a non-negligible fraction sub-mm detected sources are expected to have no counterpart below 3.6\um{} \citep[e.g.][]{Franco2018,Manning2022} and thus be \hst{} (F160W) dark.

In Figure \ref{Fig:EazY_prop} we show the photometric redshift distribution of SCUBA-2 SMG NIRCam counterparts (hereafter NIRCam-SMGs) identifying a median redshift (and 16\qth\,--\,84\qth\, quartile) of $\tilde{z}$\,=\,2.29$^{+3.58}_{-1.35}$ which is comparable to that identified in \citet{Zavala2018} for 850\um\ selected SMGs. The redshift distribution is also comparable to other studies of single-dish detected SMGs \citep[e.g.][]{Wardlow2011,Stach2019} indicating the NIRCam-SMGs do not represent a new population of higher redshift SMGs, rather a population undetected by {\em HST}. For 15 NIRCam-SMGs we also have a spectroscopic redshift from archival observations. In the inset panel of Figure \ref{Fig:EazY_prop} we show the correlation between spectroscopic and photometric redshift identifying a median ratio (and 16\qth\,--\,84\qth\, quartile) of $z_{\rm phot}$/$z_{\rm spec}$\,=\,1.0$^{+1.10}_{-0.94}$. If a spectroscopic redshift is available for the NIRCam-SMG we adopt this redshift, as opposed to the \texttt{EaZy-py} photometric redshift, in all following analysis. In future papers, we will present both the complete SED properties of the NIRCam-SMGs (Lee \& Vijayan et al. in prep.) and their close environment (Gullberg et al. in prep.). In the following section, we focus on quantifying the multi-wavelength morphology of the NIRCam-SMGs.

\begin{figure*}[h!]
    \centering
    \includegraphics[width=\linewidth]{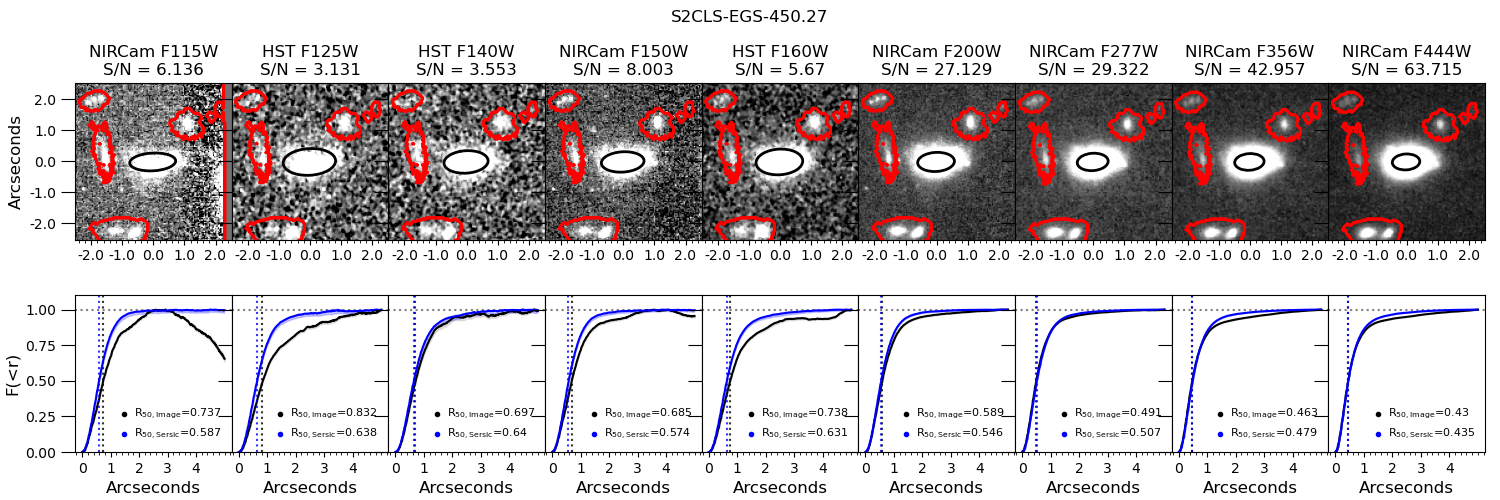}
    \caption{An example of the non-parametric growth-curve analysis for one of the NIRCam-SMGs. For each wavelength band, we show the 5 arcsecond cutout, masked regions (red contours) and derived half-light radius (black ellipse) in the upper panel. In the lower panel we show the growth curve from the image (black curve) and growth from the parametric model (blue curve) derived in Section \ref{Sec:para}.}
     \label{Fig:gc}
\end{figure*}

\section{Multi-wavelength Morphology}\label{sec:morph}
As shown in Figure~\ref{Fig:col_img}, the NIRCam-SMGs exhibit a diverse range of stellar morphology, from compact spheroidal galaxies to extended spiral structures and even potentially merging systems. As a first attempt to quantify the morphology of the galaxies, we employ a non-parametric approach, in which no prior assumptions about the galaxies' structure are made.

Throughout this section, the analysis is carried out on 0\farcs{0}4 per pixel 10 arcsecond cutouts in each filter centred on the NIRCam-SMG identified in Section \ref{Sec:counterparts}. We use the segmentation map generated from \texttt{SEP} (Section \ref{Sec:JWST}) to mask other sources in the cutout, which we dilate using the binary dilation routine in \texttt{photutils} \citep[][]{Phot2022} to ensure full masking of contaminants. We further use \texttt{photutils} to model (and remove) the background level in each cutout as well as to quantify the root mean-square (rms) noise. In the following sections `cutout' refers to this 10-arcsecond, background subtracted masked cutout and is used in the morphological analysis that follows.

Prior to measuring the morphology of the galaxies we derive the point spread function (PSF) in each NIRCam and \hst{} band. For NIRCam we first use \texttt{WebbPSF} \citep{Perrin2014} to generate PSF models across the NIRCam detectors for both short- and long-wavelength channels which are calibrated with wavefront models at the epoch of the CEERS observations. For \hst{} we use well-defined PSF models for WFC3 incorporated into \texttt{grizli}\footnote{\url{https://github.com/gbrammer/grizli-psf-library}}. The PSFs for each filter are then inserted into individual exposures of the final mosaic and drizzled to the final world coordinate system (wcs) solution. For each NIRCam-SMG we evaluate the PSF model at the position of the galaxy, resulting in a position-dependent PSF that accurately resembles the PSF of the final mosaics used in our analysis and accounts for the detector level variations. 

\subsection{Non-parametric Morphology}\label{sec:non-para}

To constrain the multi-wavelength morphology of the NIRCam-SMGs, we first derive the galaxies' non-parametric morphology via two independent methods. We first perform a curve of growth analysis in each of the NIRCam and \hst{} WFC3 wavelength bands above 1\um\ for the 43 NIRCam-SMGs. This is achieved by fitting a Gaussian profile to the cutout of each galaxy, allowing the centroid (x,y), axis ratio (b/a) and position angle (PA) to vary. We note the original centroid of the cutout is derived from the \texttt{SEP} source detection on the stacked long-wavelength NIRCam bands (Section \ref{Sec:JWST}), and thus may not be the true centre of the galaxy at shorter (bluer) wavelengths. A curve of growth is then derived in each band using ellipses which align to the galaxies axis ratio (b/a) and position angle (PA).

From the curve of growth, examples of which are shown in Figure~\ref{Fig:gc}, we measure the convolved 20, 50 and 80 per cent radii of each galaxy. The intrinsic radii of the galaxies' are derived by de-convolving the sizes with the PSF size in each band, measured through a similar curve of growth analysis. To quantify the uncertainty on the intrinsic radius of a galaxy, we perform bootstrapping over the rms noise of the cutout with 1000 iterations. The final intrinsic 20, 50 and 80 per cent radii (and uncertainty) of each galaxy are defined as the median (and standard deviation) of these 1000 iterations.

We derive a median (and 16\qth\,--\,84\qth\, quartile) axis ratio of b/a\,=\,0.67$^{+0.88}_{-0.53}$ in the NIRCam F444W band. This is comparable to the axis ratio of b/a\,$\sim$\,0.6 expected for a population of triaxial ellipsoids and higher than that expected for randomly orientated discs with exponential light profiles \citep[e.g.][]{Law2012,VW2014,Zhang2019,Robertson2022,Kart2022}. The dust (850\um) morphology of SMGs have also been shown to exhibit similar axis ratios, with \citet{Gullberg2019} reporting a median axis ratio of b/a\,=\,0.62. The median half-light radius in F444W band is R$_{\rm h}$\,=\,3.26$^{+4.66}_{-2.07}$\,kpc, which is comparable to that derived from similar studies of SMGs \citep[e.g.][]{Zavala2018,Ling2022,Cheng2023}, although we note the rest-frame wavelength probed by these studies various.

We further measure the Concentration index (\textit{C}) of the galaxies stellar light profile in each wavelength band using  Equation 4 from \citet{Conselice2014} defined as,
\begin{equation}
\rm C=5 \times \log_{10} \left( \frac{r_{outer}}{r_{inner}}\right),
\end{equation}
where \textit{C}$_{\rm 28}$ uses $r_{\rm inner}\,=\,20\%$ and  $r_{\rm outer}\,=\,80\%$. A higher Concentration indicates a larger fraction of the galaxies light is contained within the central regions. For the NIRCam-SMGs, we derive a median (and 16\qth\,--\,84\qth{} quartile) Concentration index of  \textit{C}$_{\rm 28}$\,=\,3.77$^{+4.89}_{-3.11}$ in the NIRCam F444W band. The 16\qth\,--\,84\qth{} quartile is comparable to the range derived for local late-type discs and ellipticals from optical $R$-band imaging (0.65\um) \citep{Conselice2003}. However, at the median redshift of our sample the F444W band probes the rest-frame $\sim$1\um\ emission, so this is not a direct comparison, as we analyse in Section \ref{sc:rf}.

To provide an alternative measure of half-light radius and Concentration index, as well as more quantiative morphological indicators, we use the \texttt{statmorph}\footnote{\url{https://statmorph.readthedocs.io/en/latest/}} code \citep{statmorph2019}. We run \texttt{statmorph} on the NIRCam and \hst{} WFC3 wavelength bands above 1\um\ for each of the 43 NIRCam-SMGs, using the same segmentation and PSFs as for the growth curve analysis described above. The \texttt{statmorph} code derives the Concentration, Asymmetry and Clumpiness (CAS; \citealt{Lotz2008,Conselice2014}) parameters which quantify how concentrated, asymmetrical and clumpy the galaxies' surface brightness profiles are, with higher values indicating more concentrated, asymmetric, clumpier light profiles.

In addition, the  Gini and M$_{20}$ parameters are also derived (for full definitions see \citealt{Lotz2004,Synder2015}). In short, the Gini parameter defines the pixel distribution of the galaxy's light, where G\,=\,1 means all of the light is concentrated in one pixel whilst G\,=\,0 indicates each pixel contributes equally. The M$_{20}$ parameter measures the moment of the galaxy's brightest regions containing 20 per cent of the total flux. This is then normalised by the total light moment for all pixels.  Highly negative values indicates a high concentration of light, not necessarily at the centre of the galaxy. 

To quantify the uncertainty on the non-parametric measures of morphology, we run \texttt{statmorph} 1000 times, each time adding random noise to the cutout of magnitude equal to the original rms noise measured in the cutout. The final properties (and uncertainty) for each galaxy are defined as the median (and standard deviation) of these 1000 iterations. To validate the robustness of the \texttt{statmorph} measurements we compare the half-light radius and Concentration index to that derived from our growth curve analysis. We derive a median (and 16\qth\,--\,84\qth{} quartile) growth curve to \texttt{statmorph} Concentration index ratio of  \textit{C}$_{\rm 28,GC}$\,/\,\textit{C}$_{\rm 28,statmorph}$\,=\,1.08$^{+1.42}_{-0.32}$ and median half-light radius ratio of  R$_{\rm h,GC}$\,/\,R$_{\rm h,statmorph}$\,=\,1.0$^{+1.18}_{-0.90}$, indicating good agreement between the independent morphological measurements\footnote{We note however we expect some variation between the two methods due to the definitions of centroid and total fluxes used in \texttt{statmorph} (see \citealt{Lotz2004})}. 

Having quantified the non-parametric morphology of the NIRCam-SMGs we now have multi-wavelength morphological measurements for the 43 NIRCam-SMGs from 1.15 to 4.44 \um. This sample however encapsulates a broad range of rest-frame wavelengths given the redshift range of the NIRCam-SMGs (see Figure \ref{Fig:EazY_prop}) and therefore can not be directly compared, as we discuss in the next section.

 \begin{figure*}[h!]
    \centering
    \includegraphics[width=\linewidth]{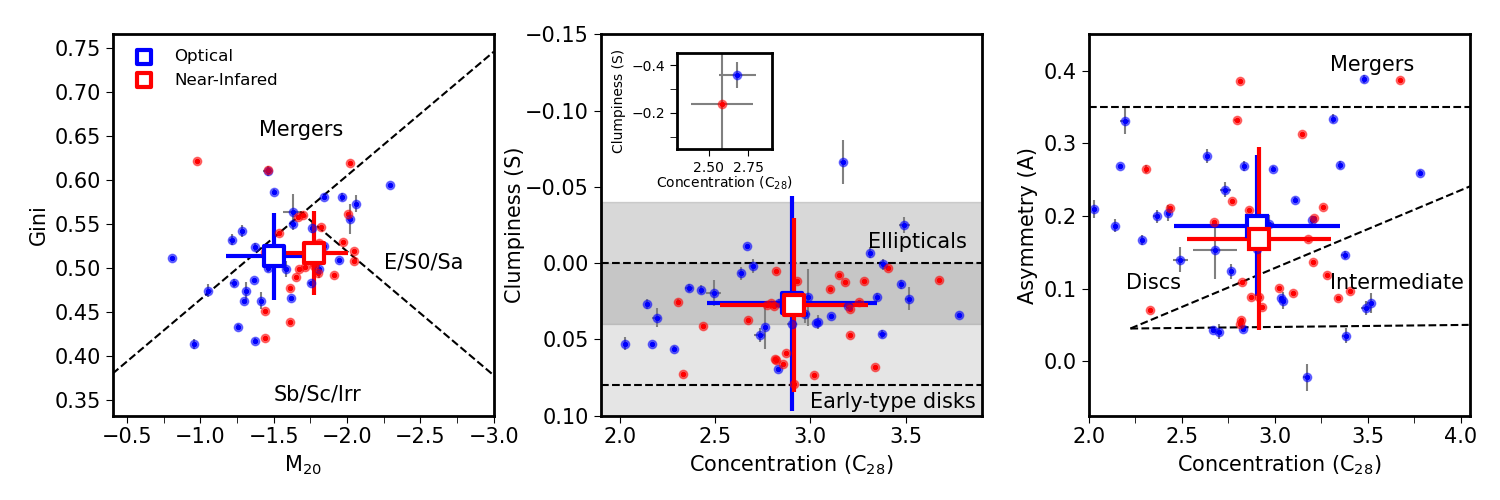}
    \caption{The non-parametric rest-frame optical and near-infrared morphology of the NIRCam-SMGs. We show the Gini and M$_{20}$ plane (\textbf{left}). The dashed black lines indicate the  division between Mergers, Elliptical and Disc galaxies as defined by \citet{Lotz2008}. In the \textbf{middle} panel we show the correlation between Clumpiness (S) and Concentration (\textit{C}$_{\rm 28}$), with the parameter space of local galaxies from \citet{Conselice2003} overlaid. Finally, in the \textbf{right} panel we show the Asymmetry ($A$) Concentration (\textit{C}$_{\rm 28}$) plane with dashed lines showing the distinction between Mergers, Disc, Intermediate and Ellipticals from \citet{Bershady2000} and \citet{Conselice2003}. In all three panels, the optical and near-infrared samples occupy the region with disc-like morphology.}
    \label{Fig:non-para}
\end{figure*}

\subsubsection{Rest-Frame Morphology}\label{sc:rf}

For each NIRCam-SMG, we determine the rest-frame wavelength probed by the multi-wavelength imaging using the galaxy's redshift, separating the wavelengths into rest-frame optical (0.25 to 0.75\um) and near-infrared (1.0 to 1.5 \um) samples. In addition, \texttt{statmorph} returns the signal to noise (S/N) per pixel for each cutout as well as a quality \texttt{flag} of the non-parametric measurements, with \texttt{flag}=0 indicating robust measurements were obtained.

If a galaxy has multiple optical or near-infrared measurements we select the measurement with the highest S/N per pixel and further remove any measurements with \texttt{statmorph} \texttt{flag}\,$\geq$\,1 and S/N per pixel\,$<$\,2.5 to ensure the most robust morphological parameters are analysed for each galaxy.
In total, this results in 32 and 26 robust non-parametric morphological measurements in the optical and near-infrared respectively for the NIRCam-SMGs for which we summarise the median properties in Table \ref{tab:morph1}. 

The Concentration and Asymmetry of the NIRCam-SMGs at both rest-frame optical and near-infrared wavelengths (as quantified in Table \ref{tab:morph1}) are comparable to that identified in local irregular galaxies from $R$-band (0.65\um) imaging of \textit{C}$_{\rm 28}$\,=\,2.7\,$\pm$\,0.2  and \textit{A}\,=\,0.17\,$\pm$\,0.11 \citep{Conselice2014}. However, the local irregular galaxies studied in \citet{Conselice2014} have much higher clumpiness (\textit{A}\,=\,0.40\,$\pm$\,0.20) than the NIRCam-SMGs. We suspect this is driven by resolution effects, smoothing out the morphology of the high-redshift galaxies.

In Figure \ref{Fig:non-para} we correlate the different non-parametric morphology measures, showing the Gini\,--\,M$_{20}$ (\textit{left}), Clumpiness (S) \,--\, Concentration (\textit{middle}) and Asymmetry \,--\, Concentration (\textit{right}) correlations for the optical and near-infrared samples. In each panel, we overlay the different morphological classifications which distinguish between mergers, disc-like galaxies and ellipticals in the local Universe \citep[e.g.][]{Lotz2008, Conselice2003, Bershady2000}. On average both optical and near-infrared measurements lie in the disc-like region of the different planes, with some scatter towards intermediate, elliptical-like systems, reflecting the range of morphology seen in Figure~\ref{Fig:col_img}. This is typical of the rest-frame optical properties of galaxies identified in the CEERS survey from $z$\,=\,3 to $z$\,=\,9 \citep{Kart2022} as well as other studies of SMGs non-parametric morphology \citep[e.g.][]{Ling2022}. The average optical measurement has a larger M$_{20}$ value at fixed Gini parameter, pushing it closer to the merger region of the Gini\,--\,M$_{20}$ diagram. We infer this could be driven by the optical measurements probing more recent events of star formation (e.g. spiral arms, star-forming clumps) as opposed to the near-infrared measurements that probe the bulk of the older stellar population. We, therefore, expect the optical morphology to be more clumpy and asymmetrical compared to the near-infrared measurements.

To investigate this further, for galaxies which have both an optical and near-infrared measurement in Figure~\ref{Fig:non-para} (26 galaxies), we quantify the ratio between the two measurements for each non-parametric parameter. On average we identify no significant offset between the optical and near-infrared Concentration or Gini parameters with median ratios of $\rm C_{\rm 28,NIR}/C_{\rm 28,OPT}$\,=\,1.02 and  $\rm Gini_{\rm NIR}/Gini_{\rm OPT}$\,=\,1.00 respectively but with an offset in the M$_{20}$ parameter of $\rm M_{\rm 20,NIR}/M_{\rm 20,OPT}$\,=\,1.1. Whilst the optical measurements, for individual galaxies, have on average larger asymmetry and clumpiness with  $\rm A_{\rm NIR}/A_{\rm OPT}$\,=\,0.95 and  $\rm S_{\rm NIR}/S_{\rm OPT}$\,=\,0.73 respectively, reflecting the fact that the optical morphologies trace recent star-formation.

There are a number of NIRCam-SMGs (9) that fall into the non-parametric merger classification in Figure~\ref{Fig:non-para}. These objects are indicated by the `M' in Figure~\ref{Fig:col_img}. Visual inspection of their morphology further indicates the presence of potential galaxy interactions and we also flag two other NIRCam-SMGs as mergers, S2CLS-EGS-450.18 and S2CLS-EGS-850.108, from visual inspection. Previous studies of SMGs have suggested they commonly reside in over-dense environments and are prone to galaxy interactions \citep[e.g.][]{Conselice2003b,Oteo2018,An2019,Torres2022,Calvi2023}. However, this is in contrast to more recent \jwst{} studies of ALMA-detected SMGs that on average identify isolated systems with stable disc-like morphologies \citep[e.g.][]{Chen2022,Cheng2023}. To investigate the NIRCam-SMGs morphology further, we undertake a parametric analysis of the non-merging systems, as discussed in the next section, using the non-parametric measures (e.g. $x$,$y$, PA, b/a) as priors.

\begin{table}[]
    \centering
    \caption{The median (and 1\,--\,$\sigma$) non-parametric morphology of the NIRCam-SMGs at rest-frame optical and near-infrared wavelengths.}
    \begin{tabular}{ccc}
    \hline
   & Optical & Near-IR  \\
 Property   &  ($0.25<\lambda<0.75\um$) & ($1.0<\lambda<1.5\um$) \\
    \hline
  \textit{C}$_{28}$ & 2.90\,$\pm$\,0.45 & 2.91\,$\pm$\,0.39\\
  \textit{A} & 0.19\,$\pm$\,0.10 & 0.17\,$\pm$\,0.13 \\
  \textit{S} &  0.03\,$\pm$\,0.07 &  0.03\,$\pm$\,0.06 \\
   Gini & 0.51\,$\pm$\,0.05 &  0.52\,$\pm$\,0.05 \\
   M$_{20}$ & $-$1.50\,$\pm$\,0.33 &  $-$1.77\,$\pm$\,0.23 \\
     \hline
    \end{tabular}
    \label{tab:morph1}
\end{table}

\subsection{Parametric Morphology}\label{Sec:para}

The parametric modelling of the SMG counterparts is carried out using \texttt{GalfitM}\footnote{\url{https://www.nottingham.ac.uk/astronomy/megamorph/}} \citep{Haubler2013}. \texttt{GalfitM} is a multi-wavelength wrapper for \texttt{Galfit} \citep{Peng2010}, in which a range of parametric models can be fit to a galaxy's morphology as a function of wavelength. The model parameters can be constrained to be independent at each wavelength provided or vary following some functional form \citep{Haubler2022}.

For our analysis, we fit\footnote{We do not perform simultaneous modelling of galaxies since we have masked out all other objects (see Section \ref{sec:morph}).} all wavelength bands above 1\um\ (with a S/N per pixel\,$>$\,2.5) from \hst{} WFC3 and \jwst{} NIRCam with a single S\'ersic model, convolved with the PSF, using \texttt{GalfitM}. We fix the centroid (x, y), axis ratio (b/a), and position angle (PA) to be constant across the wavelength bands, allowing the effective radius (R$_{\rm e}$) and S\'ersic index ($n$) to vary linearly whilst the  magnitude is modelled quadratically as a function of wavelength \citep{Haubler2022}. We also provide \texttt{GalfitM} with a sigma image for each source generated from the \texttt{ERR}\footnote{The \texttt{ERR} extension is derived from the \texttt{Grizli} pipeline and contains the read and Poisson noise of the observations.} extension of the final mosaics.

We identify a median (and 16\qth\,--\,84\qth\, quartile) ratio of \texttt{
GalfitM} to \texttt{statmorph} S\'ersic index and half-light radius of \textit{n}$_{\rm GalfitM}$\,/\,\textit{n}$_{\rm statmorph}$\,=\,0.98$^{+1.12}_{-0.76}$ and \textit{R}$_{\rm h, GalfitM}$\,/\,\textit{R}$_{\rm h, statmorph}$\,=\,0.97$^{+1.05}_{-0.81}$ respectively, indicating strong agreement between the two independent measures of parametric morphology. Despite the agreement between the two parametric codes, it does not guarantee that the galaxies' underlying morphology is well-modelled by a single S\'ersic model. Given the complex morphologies seen in Figure~\ref{Fig:col_img} and the broad scatter in non-parametric morphology in Figure~\ref{Fig:non-para}, as well as the diverse morphology identified in previous SMG studies \citep[e.g][]{Conselice2003b,Chang2018,Calvi2023}, the NIRCam-SMGs morphology may in fact be more accurately described by a two-component model.

\begin{figure*}[h!]
    \centering
    \includegraphics[width=\linewidth]{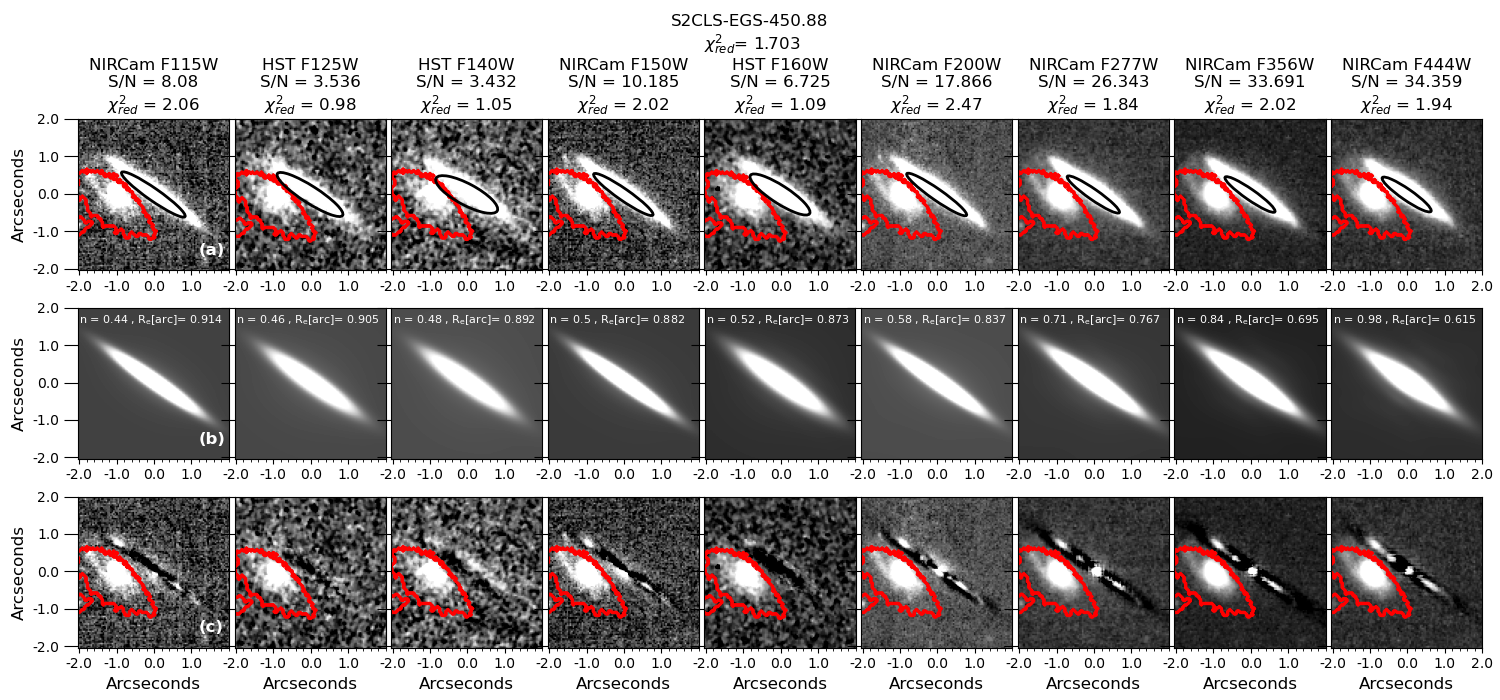}
    \caption{An example of the parametric morphological analysis for a NIRCam-SMG (S2CLS-EGS-450.88) for which we have coverage in every wavelength band. For each filter, we show a $5\arcsec$ cutout (\textbf{a}) with masked regions (red contours) and a black ellipse which indicates the growth curve derived half-light radius (Section \ref{sec:non-para}). The multi-wavelength \texttt{galfitM} model and residual are shown in panels (\textbf{b}) and (\textbf{c}) respectively with masked regions indicated as before. For each filter, we indicate the \texttt{statmorph} signal to noise per pixel as well as the reduced chi-squared per band, and overall, from \texttt{GalfitM}.}
     \label{Fig:galfit}
\end{figure*}

To investigate this further, we perform two-component multi-wavelength S\'ersic fits using \texttt{GalfitM}. We constrain the first component to be an exponential disc ($n=1$) and allow the second component's S\'ersic index to vary. This is similar to the approach taken in studies of intermediate redshift galaxies with \hst{}, whose resolved morphologies often show complex multi-component features \citep[e.g,][]{Reis2020}. We further constrain the centroid of the two components to be co-located within $\pm$3 pixels (0\farcs{12}), whilst the axis ratio, position angle and magnitude of the models are constrained in the same way as the single S\'ersic fits described above.

To assess whether the two-component fitting is more appropriate, we quantify the quality of \texttt{GalfitM} modelling using both the multi-wavelength reduced chi-squared ($\chi^{2}_{\rm red.}$) and the Bayesian Information Criteria \citep[BIC;][]{Liddle2007} which are defined as,
\begin{align}
    \chi^{2}\,&= \, \sum_{i}^{N_{\rm d}} r_{i}^{2} \\
    \chi^{2}_{\rm red.}\,&= \, \chi^{2} / (N_{\rm  d}\,- \,N_{\rm  varys}) \\
    {\rm BIC}\,&=\, \chi^{2}+N_{\rm  varys} \times \ln(N_{\rm d})
\end{align}
where $\displaystyle\sum^{N_{\rm d}}_{i}r_{i}^{2}$ is the sum of the residual image, $N_{\rm d}$ is the number of data points and $N_{\rm  varys}$ is the number of variable parameters. Both of these statistical parameters reflect the goodness of fit whilst the BIC parameter penalises models with a large number of parameters and is commonly used to determine between two parametric models \citep[e.g.][]{Head2014,Lange2016}. The model with the lower BIC values and $\chi^{2}_{\rm red.}$ close to unity is preferred.

For the single S\'ersic fitting we identify a median $\chi^{2}_{\rm red.}$\,=\,1.76 with a  16\qth\,--\,84\qth{} quartile range of $\chi^{2}_{\rm red.}$\,=\,1.44\,--\,4.41. Whilst for two components we establish a median $\chi^{2}_{\rm red.}$\,=\,1.69 with a and 16\qth\,--\,84\qth{} quartile range of $\chi^{2}_{\rm red.}$\,=\,1.38\,--\,3.60. However, on average the two-component models have a higher BIC value than the single S\'ersic models, with $< \Delta BIC>\,=\,<BIC_{\rm single}-BIC_{\rm two}>\,=-172$. This indicates that although the two-component model produces smaller residuals, it is over-fitting the morphology of the galaxies and not providing new physical information. We note also two-component models with two free S\'ersic indexes produce similar results. We, therefore, proceed with the single S\'ersic models to represent the NIRCam-SMGs parametric morphology, an example of which is shown in Figure~\ref{Fig:galfit}.

\subsubsection{Rest-Frame Morphology}

We again split the sample into rest-frame optical and near-infrared measurements as before, removing galaxies with a single-band reduced chi-squared of $\chi^{2}_{\rm red.}$\,$>3.0$ and those objects that occupy the non-parametric merger regions in Figure~\ref{Fig:non-para}. In total, this results in 25 and 14 robust parametric morphological measurements in the optical and near-infrared of the NIRCam-SMGs respectively. The image, model and residuals for each of the galaxies for this final sample are shown in Appendix \ref{App:galfit}, for which we summarise the median parametric morphology properties in Table \ref{tab:morph2}. 

On average we identify almost twice as large optical effective radius ($\sim3$\,kpc) than in the near-infrared ($\sim$1.64\,kpc). Whilst for the S\'ersic index, the optical measurements more resemble exponential discs ($n=1$) with the rest-frame near-infrared observations having higher S\'ersic index of order $n\sim2$. This is similar to the trends seen in the non-parametric morphology and further reflects the near-infrared measurements probing older, less dust-attenuated, stellar populations in the process of bulge-formation (i.e more compact), whilst rest-frame optical observations are more sensitive to recent star-formation, and consequently dust attenuation, thus seen in the extended regions of galaxies. 

\begin{figure*}[h!]
    \centering
    \includegraphics[width=\linewidth]{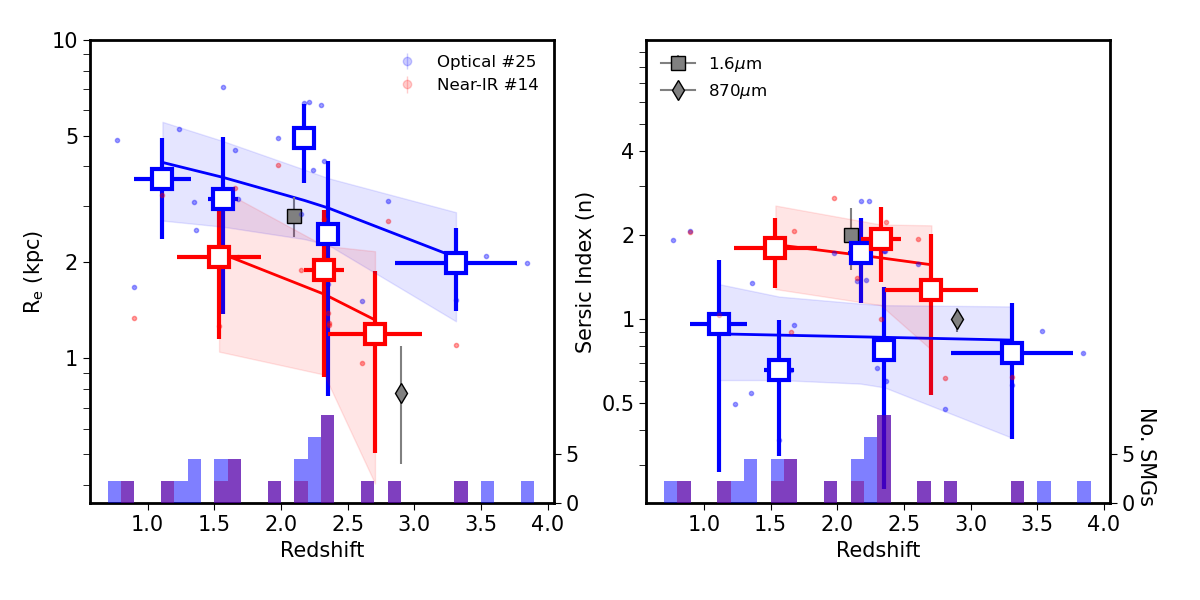}
    \caption{The median rest-frame effective radius (\re) (\textbf{left}) and  S\'ersic index ($n$) (\textbf{right}) for optical and near-infrared parametric sub-samples as a function of redshift. The squares represent the median (and 1\,--\,$\sigma$ scatter) for redshift and morphological property (\re, $n$) in a running median. On the $x$-axis we show the distribution of redshift for each optical and near-infrared sub-samples. For each sub-sample, in each panel, we show a linear parametric fit and 1-$\sigma$ scatter. We also indicate the median size and S\'ersic index from literature studies of SMGs \citep{Swinbank2014,Gullberg2019}. We identify larger (lower) effective radius (S\'ersic index) at optical wavelengths compared to near-infrared measurements, with  general a negative ($\sim 2\sigma$) correlation between the SMGs size and redshift whilst no strong correlation is identified with S\'ersic index.}
    \label{Fig:rw}
\end{figure*}

\begin{table}[]
    \centering
    \caption{The median (and 1\,--\,$\sigma$) parametric morphology of the NIRCam-SMGs at rest-frame optical and near-infrared wavelengths.}
    \begin{tabular}{lcc}
    \hline
        & Optical & Near-IR  \\
 Property   &  ($0.25<\lambda<0.75\um$) & ($1.0<\lambda<1.5\um$) \\
    \hline
   R$_{\rm e}$[kpc] & 3.10\,$\pm$\,1.67 & 1.64\,$\pm$\,0.97\\
   $n$ & 0.96\,$\pm$\,0.66 & 1.85\,$\pm$\,0.63\\
   b/a & 0.64\,$\pm$\,0.20 & 0.73\,$\pm$\,0.18\\
   $\chi^{2}_{\rm red.}$\ & 1.77\,$\pm$\,0.54 & 1.72\,$\pm$\,0.50\\
    
     \hline
    \end{tabular}
    \label{tab:morph2}
\end{table}

To analyse the cosmic evolution of the rest-frame optical and near-IR morphology of the NIRCam-SMGs, in Figure \ref{Fig:rw} we correlate the effective radius of the \texttt{GALFITM} model and the S\'ersic index  with the galaxies' redshift. We also show a running median (of 5 points) and standard deviation in each panel (squares) along with a linear parametric fit and 1-$\sigma$ uncertainty (solid line and shaded region). On each axis, we indicate the redshift distribution of the optical and near-infrared samples.

For comparison we show the median 870\um\ effective radius and S\'ersic index (\re\,$\sim$\,0.8\,kpc, $n$\,$\sim$\,1) from \citet{Gullberg2019} for 153 bright SMGs with a median redshift of $z$\,$\approx$\,2.9. We also indicate the median $H$-band (1.6\um) effective radius and S\'ersic index (\re\,$\sim$\,2.8\,kpc and $n$\,$\sim$\,2) from \citet{Swinbank2014} for 25 SMGs with a median redshift of $z$\,$\approx$\,2.1. 

At both optical and near-infrared wavelengths, we identify a negative correlation of order $\sim$2$\sigma$ between galaxy effective radius and redshift. In the optical we identify a linear slope of $\alpha$\,=\,$-$0.92\,$\pm$\,0.45 whilst for the near-infrared the linear slope is $\alpha$\,=\,$-$0.74\,$\pm$\,0.34. This evolution can be attributed to the expected growth of galaxies with cosmic time \citep[e.g.][]{VW2014a}, with the larger optical size reflecting the extended star-forming regions probed in the rest-frame optical. 

Our optical and near-infrared continuum sizes are larger than the far-infrared (dust) sizes derived in \citet{Gullberg2019}, as expected from the compact nature of the far-IR dust component of these galaxies \citep[e.g.][]{Gullberg2018,Hodge2019,Lang2019}. The rest-frame optical sizes are consistent with those reported by \citet{Zavala2018} with R$_{\rm h}$\,=\,4.8\,$\pm$\,0.4\,kpc as well as other studies of the SMG morphology with \hst{} \citep{Swinbank2014,Chen2015,Ling2022}. These studies, on average, are probing the rest-frame UV, and thus should only be compared to our optical sample. A similar offset between optical and far-IR, is found in resolved H$\alpha$ studies of $z\sim2$ SMGs \citep[e.g.][]{Chen2020} as well as in hydrodynamical simulations of typical high-redshift (1\,$<$\,$z$\,$<$\,5) main-sequence galaxies \citep[e.g.][]{Popping2022} which both identify significantly larger optical sizes compared to the galaxies' far-infrared stellar sizes. 

Using Illustris The Next Generation 50 \citep[TNG50;][]{Nelson2019} simulation, \citet{Popping2022} identify for high-redshift (2$<$\,$z$\,$<$3) massive ($\log_{10}$(M$_*$)$>$10) galaxies, a FIR (850\um) to 1.6\um{} size ratio of $\leq$\,0.5. At $z$\,=\,2.9, our linear parametric fit predicts a rest-frame optical size of 2.45\,kpc. Taking the 850\um{} size as that derived in \citet{Gullberg2019} (0.8\,kpc), we identify an 850\um{} to optical size ratio of 0.33. This is comparable to that predicted by \citet{Popping2022}, although we note this is not a direct comparison due to the different selection functions across the samples.
 
In the right-hand panel of Figure~\ref{Fig:rw} we show the S\'ersic index correlation with redshift,  identifying a linear slope of $\alpha$\,=\,$-$0.05$\pm$0.23 in the rest-frame optical whilst for the near-infrared the linear slope is $\alpha$\,=\,$-$0.25\,$\pm$\,0.44. This indicates there is no correlation with redshift for S\'ersic index both at optical and near-infrared wavelengths for the non-merging SMGs in our sample. The elevated near-infrared S\'ersic index, in comparison to the optical S\'ersic index, might be driven by the compact dusty nature of the SMGs. The dust will heavily attenuate the short-wavelength (optical) emission in the central regions of the galaxies, resulting in centrally less-peaked surface brightness profiles. As a consequence the observed size and S\'ersic index would be higher (lower) in the rest-frame optical, than for the intrinsic (un-attenuated) light profiles. 

A similar trend has been identified in recent cosmological simulations which identify that massive, intrinsically compact galaxies appear significantly more extended when the affects of dust are taken into account \citep[e.g.][]{Roper2022,Roper2023}. Our sample is intrinsically selected to have high levels of dust attenuation (Section \ref{Sec:counterparts}), and therefore dust is potentially driving the trends seen in Figure~\ref{Fig:rw}. Recent studies of high-redshift (2$<$\,$z$\,$<$6) optically-faint galaxies have identified high-dust content and compact (\re$\sim$1\,--\,2\,kpc) near-infrared morphologies \citep[e.g][]{Nelson2022}. Whilst the dominance of the disc-like morphologies in the rest-frame optical has been shown to continue out to $z$\,=\,8 \citep[e.g.][]{Leonardo2022}. These studies are in agreement with our morphological analysis of non-merging NIRCam-SMGs, with compact (extended) near-infrared (optical) sizes and disc-dominated optical morphologies out to $z$\,=\,5.4.

\section{Conclusions}\label{Sec:conc}
In this paper we have presented an analysis of the near-infrared  \jwst/NIRCam counterparts to 45 previously observed SCUBA-2 sub-mm selected galaxies (Table \ref{Table:Targets}). Through a combination of the multi-wavelength $p$-values, NIRCam colours and \texttt{Eazy-py} predicted fluxes we construct a decision tree to identify the NIRCam counterpart to each sub-mm selected SCUBA-2 source (Figure~\ref{Fig:counterparts}). We establish 43 NIRCam counterparts, from 37 SCUBA-2 sources, which cover a broad range of redshift from \zr{0.21}{5.4} with a median of $\tilde{z}$\,=\,2.29. 

We then proceed to quantify the rest-frame optical and near-infrared morphology of these 43 NIRCam-SMGs, utilising a variety of parametric and non-parametric morphological measures and taking advantage of the unprecedented depth and resolution of the CEERS \jwst{} NIRCam observations (Figure~\ref{Fig:col_img}). We establish, on average, the non-parametric measures of the NIRCam-SMGs morphology indicate late-type discs but with a broad scatter with some galaxies exhibiting galaxy interactions and mergers, whilst others resemble elliptical spheroidal systems (Figure~\ref{Fig:non-para}). For individual NIRCam-SMGs, we establish the rest-frame optical morphology is on average more asymmetrical and clumpy in comparison to near-infrared morphology, with the largest offset towards higher M$_{20}$ parameter (Table \ref{tab:morph1}). We infer this is likely driven by the rest-frame optical morphology reflecting recent and on-going star-formation whilst the near-infrared probes the older stellar populations.
 
The parametric morphology of the galaxies is quantified using \texttt{GALFITM} and is well-modelled by a single component S\'ersic model, as we identify two-component models result in over fitting of the data. On average the optical stellar continuum half-light radii exhibit larger sizes (Table \ref{tab:morph2}) than the near-infrared measurements, likely driven by dust attenuation at UV and optical wavelengths. We establish a negative correlation with effective radius and redshift (i.e larger sizes at later cosmic times) (Figure~\ref{Fig:rw}) reflecting the general growth of galaxies and build-up of stellar mass with cosmic time. The median S\'ersic index at optical wavelengths is identified to be a factor of 2 larger than at near-infrared wavelengths, with both exhibiting minimal evolution with cosmic time (Figure~\ref{Fig:rw}). The implications of dust on the observed properties of the NIRCam-SMGs are likely driving the offsets between near-infrared and optical morphologies in addition to the age of the stellar populations probed. Our results are consistent with the picture of inside-out galaxy evolution, with more centrally concentrated older stellar populations, and more extended, younger star-forming regions whose stellar emission is heavily attenuated in the central regions.
 
Overall we have shown the morphology of single-dished detected SMGs at optical and near-infrared wavelengths is diverse. ranging from compact spheroidal galaxies, grand design spirals and interacting merging systems. We have identified an evolution in morphology from small, disc-dominated structures to more extended galaxies with disc-like optical morphologies and more centrally concentrated (late-type disc) near-infrared morphologies. Future quantification of the morphology and underlying stellar population of SMGs with \jwst/MIRI imaging and spectroscopy will further build upon this picture and the wider cosmic evolution of SMGs.

\begin{acknowledgements}
      We thank Stephen Wilkins, Julie Wardlow and Oliver Newton for helpful discussions on this project. We would also like to thank the CEERS team for designing and executing the Early Release Science observations upon which this work is based. The observations analysed in this work are made with the NASA/ ESA/ CSA James Webb Space Telescope (DOI: \hyperlink{https://archive.stsci.edu/doi/resolve/resolve.html?doi=10.17909/z7p0-8481}{10.17909/z7p0-8481})
      The Cosmic Dawn Center (DAWN) is funded by the Danish National Research Foundation (DNRF) under grant No. 140. 
      BG acknowledges support from the Carlsberg Foundation Research Grant CF20-0644 ‘Physical pRoperties of the InterStellar Medium in Luminous Infrared Galaxies at High redshifT: PRISM- LIGHT’. APV, IJ \& TRG are grateful for support from the Carlsberg Foundation via grant No.~CF20-0534. GM acknowledges the Villum Fonden research grants 13160 and 37440. SJ is supported by the European Union's Horizon Europe research and innovation program under the Marie Sk\l{}odowska-Curie grant agreement No. 101060888. VK acknowledges funding from the Dutch Research Council (NWO) through the award of the Vici Grant VI.C.212.036. Cloud-based data processing and file storage for this work is provided by the AWS Cloud Credits for Research program.
\end{acknowledgements}

\hypertarget{orcids}{\section*{ORCIDs}}
Steven Gillman \orcid{0000-0001-9885-4589}\\
Bitten Gullberg \orcid{0000-0002-4671-3036},\\
Gabe Brammer \orcid{0000-0003-2680-005X},\\
Aswin P. Vijayan \orcid{0000-0002-1905-4194},\\
Minju Lee \orcid{0000-0002-2419-3068},\\
David Blánquez \orcid{0000-0001-7880-8841},\\
Malte Brinch \orcid{0000-0002-0245-6365},\\
Thomas R. Greve \orcid{0000-0002-2554-1837},\\
Iris Jermann \orcid{0000-0002-2624-1641},\\  
Shuowen Jin \orcid{0000-0002-8412-7951},\\
Vasily Kokorev \orcid{0000-0002-5588-9156},\\
Lijie Liu, \orcid{0000-0003-4956-706X}, \\
Georgios Magdis \orcid{0000-0002-4872-2294},\\
Francesca Rizzo \orcid{0000-0001-9705-2461},\\
Francesco Valentino \orcid{0000-0001-6477-4011}
       
\section*{Software}

Astropy \citep{2013A&A...558A..33A,2018AJ....156..123A},  
Photutils \citep{Phot2022}, 
Source Extractor \citep{1996A&AS..117..393B}
SEP \citep{Barbary2016}
Eazy-py \citep{Brammer2021}
GriZli \citep{Brammer2022}
GalfitM \citep{Haubler2013}

\bibliography{master}{}

\begin{thebibliography}{118}
\expandafter\ifx\csname natexlab\endcsname\relax\def\natexlab#1{#1}\fi

\bibitem[{{An} {et~al.}(2019){An}, {Simpson}, {Smail}, {Swinbank}, {Ma}, {Liu},
  {Lang}, {Schinnerer}, {Karim}, {Magnelli}, {Leslie}, {Bertoldi}, {Chen},
  {Geach}, {Matsuda}, {Stach}, {Wardlow}, {Gullberg}, {Ivison}, {Ao}, {Coogan},
  {Thomson}, {Chapman}, {Wang}, {Wang}, {Yang}, {Asquith}, {Bourne}, {Coppin},
  {Hine}, {Ho}, {Hwang}, {Kato}, {Lacaille}, {Lewis}, {Oteo}, {Scholtz},
  {Sawicki}, \& {Smith}}]{An2019}
{An}, F.~X., {Simpson}, J.~M., {Smail}, I., {et~al.} 2019, \apj, 886, 48

\bibitem[{{An} {et~al.}(2018){An}, {Stach}, {Smail}, {Swinbank}, {Almaini},
  {Simpson}, {Hartley}, {Maltby}, {Ivison}, {Arumugam}, {Wardlow}, {Cooke},
  {Gullberg}, {Thomson}, {Chen}, {Simpson}, {Geach}, {Scott}, {Dunlop},
  {Farrah}, {van der Werf}, {Blain}, {Conselice}, {Micha{\l}owski}, {Chapman},
  \& {Coppin}}]{An2018}
{An}, F.~X., {Stach}, S.~M., {Smail}, I., {et~al.} 2018, \apj, 862, 101

\bibitem[{{Astropy Collaboration} {et~al.}(2018){Astropy Collaboration},
  {Price-Whelan}, {Sip{\H{o}}cz}, {G{\"u}nther}, {Lim}, {Crawford}, {Conseil},
  {Shupe}, {Craig}, {Dencheva}, {Ginsburg}, {VanderPlas}, {Bradley},
  {P{\'e}rez-Su{\'a}rez}, {de Val-Borro}, {Aldcroft}, {Cruz}, {Robitaille},
  {Tollerud}, {Ardelean}, {Babej}, {Bach}, {Bachetti}, {Bakanov}, {Bamford},
  {Barentsen}, {Barmby}, {Baumbach}, {Berry}, {Biscani}, {Boquien}, {Bostroem},
  {Bouma}, {Brammer}, {Bray}, {Breytenbach}, {Buddelmeijer}, {Burke},
  {Calderone}, {Cano Rodr{\'\i}guez}, {Cara}, {Cardoso}, {Cheedella}, {Copin},
  {Corrales}, {Crichton}, {D'Avella}, {Deil}, {Depagne}, {Dietrich}, {Donath},
  {Droettboom}, {Earl}, {Erben}, {Fabbro}, {Ferreira}, {Finethy}, {Fox},
  {Garrison}, {Gibbons}, {Goldstein}, {Gommers}, {Greco}, {Greenfield},
  {Groener}, {Grollier}, {Hagen}, {Hirst}, {Homeier}, {Horton}, {Hosseinzadeh},
  {Hu}, {Hunkeler}, {Ivezi{\'c}}, {Jain}, {Jenness}, {Kanarek}, {Kendrew},
  {Kern}, {Kerzendorf}, {Khvalko}, {King}, {Kirkby}, {Kulkarni}, {Kumar},
  {Lee}, {Lenz}, {Littlefair}, {Ma}, {Macleod}, {Mastropietro}, {McCully},
  {Montagnac}, {Morris}, {Mueller}, {Mumford}, {Muna}, {Murphy}, {Nelson},
  {Nguyen}, {Ninan}, {N{\"o}the}, {Ogaz}, {Oh}, {Parejko}, {Parley}, {Pascual},
  {Patil}, {Patil}, {Plunkett}, {Prochaska}, {Rastogi}, {Reddy Janga},
  {Sabater}, {Sakurikar}, {Seifert}, {Sherbert}, {Sherwood-Taylor}, {Shih},
  {Sick}, {Silbiger}, {Singanamalla}, {Singer}, {Sladen}, {Sornarajah},
  {Streicher}, {Teuben}, {Thomas}, {Tremblay}, {Turner}, {Terr{\'o}n}, {van
  Kerkwijk}, {de la Vega}, {Watkins}, {Weaver}, {Whitmore}, {Woillez},
  {Zabalza}, \& {Astropy Contributors}}]{2018AJ....156..123A}
{Astropy Collaboration}, {Price-Whelan}, A.~M., {Sip{\H{o}}cz}, B.~M., {et~al.}
  2018, \aj, 156, 123

\bibitem[{{Astropy Collaboration} {et~al.}(2013){Astropy Collaboration},
  {Robitaille}, {Tollerud}, {Greenfield}, {Droettboom}, {Bray}, {Aldcroft},
  {Davis}, {Ginsburg}, {Price-Whelan}, {Kerzendorf}, {Conley}, {Crighton},
  {Barbary}, {Muna}, {Ferguson}, {Grollier}, {Parikh}, {Nair}, {Unther},
  {Deil}, {Woillez}, {Conseil}, {Kramer}, {Turner}, {Singer}, {Fox}, {Weaver},
  {Zabalza}, {Edwards}, {Azalee Bostroem}, {Burke}, {Casey}, {Crawford},
  {Dencheva}, {Ely}, {Jenness}, {Labrie}, {Lim}, {Pierfederici}, {Pontzen},
  {Ptak}, {Refsdal}, {Servillat}, \& {Streicher}}]{2013A&A...558A..33A}
{Astropy Collaboration}, {Robitaille}, T.~P., {Tollerud}, E.~J., {et~al.} 2013,
  \aap, 558, A33

\bibitem[{{Bagley} {et~al.}(2022){Bagley}, {Finkelstein}, {Koekemoer},
  {Ferguson}, {Arrabal Haro}, {Dickinson}, {Kartaltepe}, {Papovich},
  {P{\'e}rez-Gonz{\'a}lez}, {Pirzkal}, {Somerville}, {Willmer}, {Yang}, {Yung},
  {Fontana}, {Grazian}, {Grogin}, {Hirschmann}, {Kewley}, {Kirkpatrick},
  {Kocevski}, {Lotz}, {Medrano}, {Morales}, {Pentericci}, {Ravindranath},
  {Trump}, {Wilkins}, {Calabr{\`o}}, {Cooper}, {Costantin}, {de la Vega},
  {Hutchison}, {Lucas}, {McGrath}, {Wang}, \& {Wuyts}}]{Bagley2022}
{Bagley}, M.~B., {Finkelstein}, S.~L., {Koekemoer}, A.~M., {et~al.} 2022, arXiv
  e-prints, arXiv:2211.02495

\bibitem[{{Barbary} {et~al.}(2016){Barbary}, {Boone}, {McCully}, {Craig},
  {Deil}, \& {Rose}}]{Barbary2016}
{Barbary}, K., {Boone}, K., {McCully}, C., {et~al.} 2016, {Kbarbary/Sep:
  V1.0.0}, Zenodo

\bibitem[{{Barger} {et~al.}(2014){Barger}, {Cowie}, {Chen}, {Owen}, {Wang},
  {Casey}, {Lee}, {Sanders}, \& {Williams}}]{Barger2014}
{Barger}, A.~J., {Cowie}, L.~L., {Chen}, C.~C., {et~al.} 2014, \apj, 784, 9

\bibitem[{{Baugh} {et~al.}(2005){Baugh}, {Lacey}, {Frenk}, {Granato}, {Silva},
  {Bressan}, {Benson}, \& {Cole}}]{Baugh2005}
{Baugh}, C.~M., {Lacey}, C.~G., {Frenk}, C.~S., {et~al.} 2005, \mnras, 356,
  1191

\bibitem[{{Bershady} {et~al.}(2000){Bershady}, {Jangren}, \&
  {Conselice}}]{Bershady2000}
{Bershady}, M.~A., {Jangren}, A., \& {Conselice}, C.~J. 2000, \aj, 119, 2645

\bibitem[{{Bertin} \& {Arnouts}(1996{\natexlab{a}})}]{Bertin1996}
{Bertin}, E. \& {Arnouts}, S. 1996{\natexlab{a}}, \aaps, 117, 393

\bibitem[{{Bertin} \& {Arnouts}(1996{\natexlab{b}})}]{1996A&AS..117..393B}
{Bertin}, E. \& {Arnouts}, S. 1996{\natexlab{b}}, \aaps, 117, 393

\bibitem[{{Biggs} {et~al.}(2011){Biggs}, {Ivison}, {Ibar}, {Wardlow},
  {Dannerbauer}, {Smail}, {Walter}, {Wei{\ss}}, {Chapman}, {Coppin}, {De
  Breuck}, {Dickinson}, {Knudsen}, {Mainieri}, {Menten}, \&
  {Papovich}}]{Biggs2011}
{Biggs}, A.~D., {Ivison}, R.~J., {Ibar}, E., {et~al.} 2011, \mnras, 413, 2314

\bibitem[{{Blain} {et~al.}(2002){Blain}, {Smail}, {Ivison}, {Kneib}, \&
  {Frayer}}]{Blain2002}
{Blain}, A.~W., {Smail}, I., {Ivison}, R.~J., {Kneib}, J.~P., \& {Frayer},
  D.~T. 2002, \physrep, 369, 111

\bibitem[{{Boyer} {et~al.}(2022){Boyer}, {Anderson}, {Gennaro}, {Geha},
  {Wingfield McQuinn}, {Tollerud}, {Correnti}, {Brenner Newman}, {Cohen},
  {Kallivayalil}, {Beaton}, {Cole}, {Dolphin}, {Kalirai}, {Sandstrom},
  {Savino}, {Skillman}, {Weisz}, \& {Williams}}]{Boyer2022}
{Boyer}, M.~L., {Anderson}, J., {Gennaro}, M., {et~al.} 2022, Research Notes of
  the American Astronomical Society, 6, 191

\bibitem[{{Bradley} {et~al.}(2022{\natexlab{a}}){Bradley}, {Sip{\H{o}}cz},
  {Robitaille}, {Tollerud}, {Vin{\'\i}cius}, {Deil}, {Barbary}, {Wilson},
  {Busko}, {Donath}, {G{\"u}nther}, {Cara}, {Lim}, {Me{\ss}linger}, {Conseil},
  {Bostroem}, {Droettboom}, {Bray}, {Andersen Bratholm}, {Barentsen}, {Craig},
  {Rathi}, {Pascual}, {Perren}, {Georgiev}, {De Val-Borro}, {Kerzendorf},
  {Bach}, {Quint}, \& {Souchereau}}]{Phot2022}
{Bradley}, L., {Sip{\H{o}}cz}, B., {Robitaille}, T., {et~al.}
  2022{\natexlab{a}}, {astropy/photutils: 1.5.0}, Zenodo

\bibitem[{{Bradley} {et~al.}(2022{\natexlab{b}}){Bradley}, {Coe}, {Brammer},
  {Furtak}, {Larson}, {Andrade-Santos}, {Bhatawdekar}, {Bradac}, {Broadhurst},
  {Carnall}, {Conselice}, {Diego}, {Frye}, {Fujimoto}, {Y. -Y Hsiao},
  {Hutchison}, {Jung}, {Mahler}, {McCandliss}, {Oguri}, {Postman}, {Sharon},
  {Trenti}, {Vanzella}, {Welch}, {Windhorst}, \& {Zitrin}}]{Bradley2022}
{Bradley}, L.~D., {Coe}, D., {Brammer}, G., {et~al.} 2022{\natexlab{b}}, arXiv
  e-prints, arXiv:2210.01777

\bibitem[{{Brammer} \& {Matharu}(2021)}]{Brammer2021}
{Brammer}, G. \& {Matharu}, J. 2021, {gbrammer/grizli: Release 2021}, Zenodo

\bibitem[{{Brammer} {et~al.}(2022){Brammer}, {Strait}, {Matharu}, \&
  {Momcheva}}]{Brammer2022}
{Brammer}, G., {Strait}, V., {Matharu}, J., \& {Momcheva}, I. 2022, {grizli},
  Zenodo

\bibitem[{{Brammer} {et~al.}(2008){Brammer}, {van Dokkum}, \&
  {Coppi}}]{Brammer2008}
{Brammer}, G.~B., {van Dokkum}, P.~G., \& {Coppi}, P. 2008, \apj, 686, 1503

\bibitem[{{Calvi} {et~al.}(2023){Calvi}, {Castignani}, \&
  {Dannerbauer}}]{Calvi2023}
{Calvi}, R., {Castignani}, G., \& {Dannerbauer}, H. 2023, arXiv e-prints,
  arXiv:2302.10323

\bibitem[{{Cardona-Torres} {et~al.}(2022){Cardona-Torres}, {Aretxaga},
  {Monta{\~n}a}, {Zavala}, \& {Faber}}]{Torres2022}
{Cardona-Torres}, L., {Aretxaga}, I., {Monta{\~n}a}, A., {Zavala}, J.~A., \&
  {Faber}, S.~M. 2022, \mnras [\eprint[arXiv]{2210.04437}]

\bibitem[{{Casey} {et~al.}(2014){Casey}, {Narayanan}, \& {Cooray}}]{Casey2014}
{Casey}, C.~M., {Narayanan}, D., \& {Cooray}, A. 2014, \physrep, 541, 45

\bibitem[{{Chabrier}(2003)}]{Chabrier2003}
{Chabrier}, G. 2003, \pasp, 115, 763

\bibitem[{{Chang} {et~al.}(2018){Chang}, {Ferraro}, {Wang}, {Lim}, {Toba},
  {An}, {Chen}, {Smail}, {Shim}, {Ao}, {Bunker}, {Conselice}, {Cowley}, {da
  Cunha}, {Fan}, {Goto}, {Guo}, {Ho}, {Hwang}, {Lee}, {Lee}, {Micha{\l}owski},
  {Oteo}, {Scott}, {Serjeant}, {Shu}, {Simpson}, \& {Urquhart}}]{Chang2018}
{Chang}, Y.-Y., {Ferraro}, N., {Wang}, W.-H., {et~al.} 2018, \apj, 865, 103

\bibitem[{{Chapman} {et~al.}(2005){Chapman}, {Blain}, {Smail}, \&
  {Ivison}}]{Chapman2005}
{Chapman}, S.~C., {Blain}, A.~W., {Smail}, I., \& {Ivison}, R.~J. 2005, \apj,
  622, 772

\bibitem[{{Chen} {et~al.}(2022){Chen}, {Gao}, {Hsu}, {Liao}, {Ling}, {Lo},
  {Smail}, {Wang}, \& {Wang}}]{Chen2022}
{Chen}, C.-C., {Gao}, Z.-K., {Hsu}, Q.-N., {et~al.} 2022, \apjl, 939, L7

\bibitem[{{Chen} {et~al.}(2020){Chen}, {Harrison}, {Smail}, {Swinbank},
  {Turner}, {Wardlow}, {Brandt}, {Calistro Rivera}, {Chapman}, {Cooke},
  {Dannerbauer}, {Dunlop}, {Farrah}, {Micha{\l}owski}, {Schinnerer}, {Simpson},
  {Thomson}, \& {van der Werf}}]{Chen2020}
{Chen}, C.-C., {Harrison}, C.~M., {Smail}, I., {et~al.} 2020, \aap, 635, A119

\bibitem[{{Chen} {et~al.}(2016){Chen}, {Smail}, {Swinbank}, {Simpson},
  {Almaini}, {Conselice}, {Hartley}, {Mortlock}, {Simpson}, \&
  {Wilkinson}}]{Chen2016}
{Chen}, C.-C., {Smail}, I., {Swinbank}, A.~M., {et~al.} 2016, \apj, 831, 91

\bibitem[{{Chen} {et~al.}(2015){Chen}, {Smail}, {Swinbank}, {Simpson}, {Ma},
  {Alexander}, {Biggs}, {Brandt}, {Chapman}, {Coppin}, {Danielson},
  {Dannerbauer}, {Edge}, {Greve}, {Ivison}, {Karim}, {Menten}, {Schinnerer},
  {Walter}, {Wardlow}, {Wei{\ss}}, \& {van der Werf}}]{Chen2015}
{Chen}, C.-C., {Smail}, I., {Swinbank}, A.~M., {et~al.} 2015, \apj, 799, 194

\bibitem[{{Cheng} {et~al.}(2023){Cheng}, {Huang}, {Smail}, {Yan}, {Cohen},
  {Jansen}, {Windhorst}, {Ma}, {Koekemoer}, {Willmer}, {Willner}, {Diego},
  {Frye}, {Conselice}, {Ferreira}, {Petric}, {Yun}, {Gim}, {Polletta},
  {Duncan}, {Holwerda}, {R{\"o}ttgering}, {Honor}, {Hathi}, {Kamieneski},
  {Adams}, {Coe}, {Broadhurst}, {Summers}, {Tompkins}, {Driver}, {Grogin},
  {Marshall}, {Pirzkal}, {Robotham}, \& {Ryan}}]{Cheng2023}
{Cheng}, C., {Huang}, J.-S., {Smail}, I., {et~al.} 2023, \apjl, 942, L19

\bibitem[{{Cochrane} {et~al.}(2021){Cochrane}, {Best}, {Smail}, {Ibar},
  {Cheng}, {Swinbank}, {Molina}, {Sobral}, \&
  {Dudzevi{\v{c}}i{\={u}}t{\.{e}}}}]{Cochrane2021}
{Cochrane}, R.~K., {Best}, P.~N., {Smail}, I., {et~al.} 2021, \mnras, 503, 2622

\bibitem[{{Cochrane} {et~al.}(2022){Cochrane}, {Hayward},
  {Angl{\'e}s-Alc{\'a}zar}, \& {Somerville}}]{Cochrane2022}
{Cochrane}, R.~K., {Hayward}, C.~C., {Angl{\'e}s-Alc{\'a}zar}, D., \&
  {Somerville}, R.~S. 2022, \mnras [\eprint[arXiv]{2211.11702}]

\bibitem[{{Conroy} \& {Gunn}(2010)}]{Conroy2010}
{Conroy}, C. \& {Gunn}, J.~E. 2010, \apj, 712, 833

\bibitem[{{Conselice}(2014)}]{Conselice2014}
{Conselice}, C.~J. 2014, \araa, 52, 291

\bibitem[{{Conselice} {et~al.}(2003{\natexlab{a}}){Conselice}, {Chapman}, \&
  {Windhorst}}]{Conselice2003}
{Conselice}, C.~J., {Chapman}, S.~C., \& {Windhorst}, R.~A. 2003{\natexlab{a}},
  \apjl, 596, L5

\bibitem[{{Conselice} {et~al.}(2003{\natexlab{b}}){Conselice}, {Chapman}, \&
  {Windhorst}}]{Conselice2003b}
{Conselice}, C.~J., {Chapman}, S.~C., \& {Windhorst}, R.~A. 2003{\natexlab{b}},
  \apjl, 596, L5

\bibitem[{{Cowley} {et~al.}(2019){Cowley}, {Lacey}, {Baugh}, {Cole}, {Frenk},
  \& {Lagos}}]{Cowley2019}
{Cowley}, W.~I., {Lacey}, C.~G., {Baugh}, C.~M., {et~al.} 2019, \mnras, 487,
  3082

\bibitem[{{Dav{\'e}} {et~al.}(2010){Dav{\'e}}, {Finlator}, {Oppenheimer},
  {Fardal}, {Katz}, {Kere{\v{s}}}, \& {Weinberg}}]{Dave2010}
{Dav{\'e}}, R., {Finlator}, K., {Oppenheimer}, B.~D., {et~al.} 2010, \mnras,
  404, 1355

\bibitem[{{Davis} {et~al.}(2007){Davis}, {Guhathakurta}, {Konidaris}, {Newman},
  {Ashby}, {Biggs}, {Barmby}, {Bundy}, {Chapman}, {Coil}, {Conselice},
  {Cooper}, {Croton}, {Eisenhardt}, {Ellis}, {Faber}, {Fang}, {Fazio},
  {Georgakakis}, {Gerke}, {Goss}, {Gwyn}, {Harker}, {Hopkins}, {Huang},
  {Ivison}, {Kassin}, {Kirby}, {Koekemoer}, {Koo}, {Laird}, {Le Floc'h}, {Lin},
  {Lotz}, {Marshall}, {Martin}, {Metevier}, {Moustakas}, {Nandra}, {Noeske},
  {Papovich}, {Phillips}, {Rich}, {Rieke}, {Rigopoulou}, {Salim},
  {Schiminovich}, {Simard}, {Smail}, {Small}, {Weiner}, {Willmer}, {Willner},
  {Wilson}, {Wright}, \& {Yan}}]{Davis2007}
{Davis}, M., {Guhathakurta}, P., {Konidaris}, N.~P., {et~al.} 2007, \apjl, 660,
  L1

\bibitem[{{Dempsey} {et~al.}(2013){Dempsey}, {Friberg}, {Jenness}, {Tilanus},
  {Thomas}, {Holland}, {Bintley}, {Berry}, {Chapin}, {Chrysostomou}, {Davis},
  {Gibb}, {Parsons}, \& {Robson}}]{Dempsey2013}
{Dempsey}, J.~T., {Friberg}, P., {Jenness}, T., {et~al.} 2013, \mnras, 430,
  2534

\bibitem[{{Dey} {et~al.}(1999){Dey}, {Graham}, {Ivison}, {Smail}, {Wright}, \&
  {Liu}}]{Dey1999}
{Dey}, A., {Graham}, J.~R., {Ivison}, R.~J., {et~al.} 1999, \apj, 519, 610

\bibitem[{{D{\'\i}az-S{\'a}nchez} {et~al.}(2017){D{\'\i}az-S{\'a}nchez},
  {Iglesias-Groth}, {Rebolo}, \& {Dannerbauer}}]{Diaz2017}
{D{\'\i}az-S{\'a}nchez}, A., {Iglesias-Groth}, S., {Rebolo}, R., \&
  {Dannerbauer}, H. 2017, \apjl, 843, L22

\bibitem[{{Dickinson} \& {FIDEL Team}(2007)}]{Dickinson2007}
{Dickinson}, M. \& {FIDEL Team}. 2007, in American Astronomical Society Meeting
  Abstracts, Vol. 211, American Astronomical Society Meeting Abstracts, 52.16

\bibitem[{{dos Reis} {et~al.}(2020){dos Reis}, {Buitrago}, {Papaderos},
  {Matute}, {Afonso}, {Amarantidis}, {Breda}, {Gomes}, {Humphrey}, {Lobo},
  {Lorenzoni}, {Pappalardo}, {Paulino-Afonso}, \& {Scott}}]{Reis2020}
{dos Reis}, S.~N., {Buitrago}, F., {Papaderos}, P., {et~al.} 2020, \aap, 634,
  A11

\bibitem[{{Downes} {et~al.}(1986){Downes}, {Peacock}, {Savage}, \&
  {Carrie}}]{Downes1986}
{Downes}, A.~J.~B., {Peacock}, J.~A., {Savage}, A., \& {Carrie}, D.~R. 1986,
  \mnras, 218, 31

\bibitem[{{Eales} {et~al.}(1999){Eales}, {Lilly}, {Gear}, {Dunne}, {Bond},
  {Hammer}, {Le F{\`e}vre}, \& {Crampton}}]{Eales1999}
{Eales}, S., {Lilly}, S., {Gear}, W., {et~al.} 1999, \apj, 515, 518

\bibitem[{{Fazio} {et~al.}(2004){Fazio}, {Hora}, {Allen}, {Ashby}, {Barmby},
  {Deutsch}, {Huang}, {Kleiner}, {Marengo}, {Megeath}, {Melnick}, {Pahre},
  {Patten}, {Polizotti}, {Smith}, {Taylor}, {Wang}, {Willner}, {Hoffmann},
  {Pipher}, {Forrest}, {McMurty}, {McCreight}, {McKelvey}, {McMurray}, {Koch},
  {Moseley}, {Arendt}, {Mentzell}, {Marx}, {Losch}, {Mayman}, {Eichhorn},
  {Krebs}, {Jhabvala}, {Gezari}, {Fixsen}, {Flores}, {Shakoorzadeh}, {Jungo},
  {Hakun}, {Workman}, {Karpati}, {Kichak}, {Whitley}, {Mann}, {Tollestrup},
  {Eisenhardt}, {Stern}, {Gorjian}, {Bhattacharya}, {Carey}, {Nelson},
  {Glaccum}, {Lacy}, {Lowrance}, {Laine}, {Reach}, {Stauffer}, {Surace},
  {Wilson}, {Wright}, {Hoffman}, {Domingo}, \& {Cohen}}]{Fazio2004}
{Fazio}, G.~G., {Hora}, J.~L., {Allen}, L.~E., {et~al.} 2004, \apjs, 154, 10

\bibitem[{{Ferreira} {et~al.}(2022){Ferreira}, {Conselice}, {Sazonova},
  {Ferrari}, {Caruana}, {Tohill}, {Lucatelli}, {Adams}, {Irodotou}, {Marshall},
  {Roper}, {Lovell}, {Verma}, {Austin}, {Trussler}, \&
  {Wilkins}}]{Leonardo2022}
{Ferreira}, L., {Conselice}, C.~J., {Sazonova}, E., {et~al.} 2022, arXiv
  e-prints, arXiv:2210.01110

\bibitem[{{Franco} {et~al.}(2018){Franco}, {Elbaz}, {B{\'e}thermin},
  {Magnelli}, {Schreiber}, {Ciesla}, {Dickinson}, {Nagar}, {Silverman},
  {Daddi}, {Alexander}, {Wang}, {Pannella}, {Le Floc'h}, {Pope}, {Giavalisco},
  {Maury}, {Bournaud}, {Chary}, {Demarco}, {Ferguson}, {Finkelstein}, {Inami},
  {Iono}, {Juneau}, {Lagache}, {Leiton}, {Lin}, {Magdis}, {Messias},
  {Motohara}, {Mullaney}, {Okumura}, {Papovich}, {Pforr}, {Rujopakarn},
  {Sargent}, {Shu}, \& {Zhou}}]{Franco2018}
{Franco}, M., {Elbaz}, D., {B{\'e}thermin}, M., {et~al.} 2018, \aap, 620, A152

\bibitem[{{Fruchter} \& {Hook}(2002)}]{Fruchter2002}
{Fruchter}, A.~S. \& {Hook}, R.~N. 2002, \pasp, 114, 144

\bibitem[{{Gaia Collaboration} {et~al.}(2021){Gaia Collaboration}, {Brown},
  {Vallenari}, {Prusti}, {de Bruijne}, {Babusiaux}, {Biermann}, {Creevey},
  {Evans}, {Eyer}, {Hutton}, {Jansen}, {Jordi}, {Klioner}, {Lammers},
  {Lindegren}, {Luri}, {Mignard}, {Panem}, {Pourbaix}, {Randich}, {Sartoretti},
  {Soubiran}, {Walton}, {Arenou}, {Bailer-Jones}, {Bastian}, {Cropper},
  {Drimmel}, {Katz}, {Lattanzi}, {van Leeuwen}, {Bakker}, {Cacciari},
  {Casta{\~n}eda}, {De Angeli}, {Ducourant}, {Fabricius}, {Fouesneau},
  {Fr{\'e}mat}, {Guerra}, {Guerrier}, {Guiraud}, {Jean-Antoine Piccolo},
  {Masana}, {Messineo}, {Mowlavi}, {Nicolas}, {Nienartowicz}, {Pailler},
  {Panuzzo}, {Riclet}, {Roux}, {Seabroke}, {Sordo}, {Tanga}, {Th{\'e}venin},
  {Gracia-Abril}, {Portell}, {Teyssier}, {Altmann}, {Andrae}, {Bellas-Velidis},
  {Benson}, {Berthier}, {Blomme}, {Brugaletta}, {Burgess}, {Busso}, {Carry},
  {Cellino}, {Cheek}, {Clementini}, {Damerdji}, {Davidson}, {Delchambre},
  {Dell'Oro}, {Fern{\'a}ndez-Hern{\'a}ndez}, {Galluccio}, {Garc{\'\i}a-Lario},
  {Garcia-Reinaldos}, {Gonz{\'a}lez-N{\'u}{\~n}ez}, {Gosset}, {Haigron},
  {Halbwachs}, {Hambly}, {Harrison}, {Hatzidimitriou}, {Heiter},
  {Hern{\'a}ndez}, {Hestroffer}, {Hodgkin}, {Holl}, {Jan{\ss}en}, {Jevardat de
  Fombelle}, {Jordan}, {Krone-Martins}, {Lanzafame}, {L{\"o}ffler}, {Lorca},
  {Manteiga}, {Marchal}, {Marrese}, {Moitinho}, {Mora}, {Muinonen}, {Osborne},
  {Pancino}, {Pauwels}, {Petit}, {Recio-Blanco}, {Richards}, {Riello},
  {Rimoldini}, {Robin}, {Roegiers}, {Rybizki}, {Sarro}, {Siopis}, {Smith},
  {Sozzetti}, {Ulla}, {Utrilla}, {van Leeuwen}, {van Reeven}, {Abbas}, {Abreu
  Aramburu}, {Accart}, {Aerts}, {Aguado}, {Ajaj}, {Altavilla}, {{\'A}lvarez},
  {{\'A}lvarez Cid-Fuentes}, {Alves}, {Anderson}, {Anglada Varela}, {Antoja},
  {Audard}, {Baines}, {Baker}, {Balaguer-N{\'u}{\~n}ez}, {Balbinot}, {Balog},
  {Barache}, {Barbato}, {Barros}, {Barstow}, {Bartolom{\'e}}, {Bassilana},
  {Bauchet}, {Baudesson-Stella}, {Becciani}, {Bellazzini}, {Bernet}, {Bertone},
  {Bianchi}, {Blanco-Cuaresma}, {Boch}, {Bombrun}, {Bossini}, {Bouquillon},
  {Bragaglia}, {Bramante}, {Breedt}, {Bressan}, {Brouillet}, {Bucciarelli},
  {Burlacu}, {Busonero}, {Butkevich}, {Buzzi}, {Caffau}, {Cancelliere},
  {C{\'a}novas}, {Cantat-Gaudin}, {Carballo}, {Carlucci}, {Carnerero},
  {Carrasco}, {Casamiquela}, {Castellani}, {Castro-Ginard}, {Castro Sampol},
  {Chaoul}, {Charlot}, {Chemin}, {Chiavassa}, {Cioni}, {Comoretto}, {Cooper},
  {Cornez}, {Cowell}, {Crifo}, {Crosta}, {Crowley}, {Dafonte}, {Dapergolas},
  {David}, {David}, {de Laverny}, {De Luise}, {De March}, {De Ridder}, {de
  Souza}, {de Teodoro}, {de Torres}, {del Peloso}, {del Pozo}, {Delbo},
  {Delgado}, {Delgado}, {Delisle}, {Di Matteo}, {Diakite}, {Diener},
  {Distefano}, {Dolding}, {Eappachen}, {Edvardsson}, {Enke}, {Esquej}, {Fabre},
  {Fabrizio}, {Faigler}, {Fedorets}, {Fernique}, {Fienga}, {Figueras},
  {Fouron}, {Fragkoudi}, {Fraile}, {Franke}, {Gai}, {Garabato},
  {Garcia-Gutierrez}, {Garc{\'\i}a-Torres}, {Garofalo}, {Gavras}, {Gerlach},
  {Geyer}, {Giacobbe}, {Gilmore}, {Girona}, {Giuffrida}, {Gomel}, {Gomez},
  {Gonzalez-Santamaria}, {Gonz{\'a}lez-Vidal}, {Granvik},
  {Guti{\'e}rrez-S{\'a}nchez}, {Guy}, {Hauser}, {Haywood}, {Helmi}, {Hidalgo},
  {Hilger}, {H{\l}adczuk}, {Hobbs}, {Holland}, {Huckle}, {Jasniewicz},
  {Jonker}, {Juaristi Campillo}, {Julbe}, {Karbevska}, {Kervella}, {Khanna},
  {Kochoska}, {Kontizas}, {Kordopatis}, {Korn}, {Kostrzewa-Rutkowska},
  {Kruszy{\'n}ska}, {Lambert}, {Lanza}, {Lasne}, {Le Campion}, {Le Fustec},
  {Lebreton}, {Lebzelter}, {Leccia}, {Leclerc}, {Lecoeur-Taibi}, {Liao},
  {Licata}, {Lindstr{\o}m}, {Lister}, {Livanou}, {Lobel}, {Madrero Pardo},
  {Managau}, {Mann}, {Marchant}, {Marconi}, {Marcos Santos}, {Marinoni},
  {Marocco}, {Marshall}, {Martin Polo}, {Mart{\'\i}n-Fleitas}, {Masip},
  {Massari}, {Mastrobuono-Battisti}, {Mazeh}, {McMillan}, {Messina},
  {Michalik}, {Millar}, {Mints}, {Molina}, {Molinaro}, {Moln{\'a}r},
  {Montegriffo}, {Mor}, {Morbidelli}, {Morel}, {Morris}, {Mulone}, {Munoz},
  {Muraveva}, {Murphy}, {Musella}, {Noval}, {Ord{\'e}novic}, {Orr{\`u}},
  {Osinde}, {Pagani}, {Pagano}, {Palaversa}, {Palicio}, {Panahi}, {Pawlak},
  {Pe{\~n}alosa Esteller}, {Penttil{\"a}}, {Piersimoni}, {Pineau}, {Plachy},
  {Plum}, {Poggio}, {Poretti}, {Poujoulet}, {Pr{\v{s}}a}, {Pulone}, {Racero},
  {Ragaini}, {Rainer}, {Raiteri}, {Rambaux}, {Ramos}, {Ramos-Lerate}, {Re
  Fiorentin}, {Regibo}, {Reyl{\'e}}, {Ripepi}, {Riva}, {Rixon}, {Robichon},
  {Robin}, {Roelens}, {Rohrbasser}, {Romero-G{\'o}mez}, {Rowell}, {Royer},
  {Rybicki}, {Sadowski}, {Sagrist{\`a} Sell{\'e}s}, {Sahlmann}, {Salgado},
  {Salguero}, {Samaras}, {Sanchez Gimenez}, {Sanna}, {Santove{\~n}a},
  {Sarasso}, {Schultheis}, {Sciacca}, {Segol}, {Segovia}, {S{\'e}gransan},
  {Semeux}, {Shahaf}, {Siddiqui}, {Siebert}, {Siltala}, {Slezak}, {Smart},
  {Solano}, {Solitro}, {Souami}, {Souchay}, {Spagna}, {Spoto}, {Steele},
  {Steidelm{\"u}ller}, {Stephenson}, {S{\"u}veges}, {Szabados}, {Szegedi-Elek},
  {Taris}, {Tauran}, {Taylor}, {Teixeira}, {Thuillot}, {Tonello}, {Torra},
  {Torra}, {Turon}, {Unger}, {Vaillant}, {van Dillen}, {Vanel}, {Vecchiato},
  {Viala}, {Vicente}, {Voutsinas}, {Weiler}, {Wevers}, {Wyrzykowski}, {Yoldas},
  {Yvard}, {Zhao}, {Zorec}, {Zucker}, {Zurbach}, \& {Zwitter}}]{Gaia2021}
{Gaia Collaboration}, {Brown}, A.~G.~A., {Vallenari}, A., {et~al.} 2021, \aap,
  649, A1

\bibitem[{{Geach} {et~al.}(2017){Geach}, {Dunlop}, {Halpern}, {Smail}, {van der
  Werf}, {Alexander}, {Almaini}, {Aretxaga}, {Arumugam}, {Asboth}, {Banerji},
  {Beanlands}, {Best}, {Blain}, {Birkinshaw}, {Chapin}, {Chapman}, {Chen},
  {Chrysostomou}, {Clarke}, {Clements}, {Conselice}, {Coppin}, {Cowley},
  {Danielson}, {Eales}, {Edge}, {Farrah}, {Gibb}, {Harrison}, {Hine}, {Hughes},
  {Ivison}, {Jarvis}, {Jenness}, {Jones}, {Karim}, {Koprowski}, {Knudsen},
  {Lacey}, {Mackenzie}, {Marsden}, {McAlpine}, {McMahon}, {Meijerink},
  {Micha{\l}owski}, {Oliver}, {Page}, {Peacock}, {Rigopoulou}, {Robson},
  {Roseboom}, {Rotermund}, {Scott}, {Serjeant}, {Simpson}, {Simpson}, {Smith},
  {Spaans}, {Stanley}, {Stevens}, {Swinbank}, {Targett}, {Thomson}, {Valiante},
  {Wake}, {Webb}, {Willott}, {Zavala}, \& {Zemcov}}]{Geach2017}
{Geach}, J.~E., {Dunlop}, J.~S., {Halpern}, M., {et~al.} 2017, \mnras, 465,
  1789

\bibitem[{{G{\'o}mez-Guijarro} {et~al.}(2018){G{\'o}mez-Guijarro}, {Toft},
  {Karim}, {Magnelli}, {Magdis}, {Jim{\'e}nez-Andrade}, {Capak}, {Fraternali},
  {Fujimoto}, {Riechers}, {Schinnerer}, {Smol{\v{c}}i{\'c}}, {Aravena},
  {Bertoldi}, {Cortzen}, {Hasinger}, {Hu}, {Jones}, {Koekemoer}, {Lee},
  {McCracken}, {Micha{\l}owski}, {Navarrete}, {Povi{\'c}}, {Puglisi},
  {Romano-D{\'\i}az}, {Sheth}, {Silverman}, {Staguhn}, {Steinhardt},
  {Stockmann}, {Tanaka}, {Valentino}, {van Kampen}, \& {Zirm}}]{Gomez2018}
{G{\'o}mez-Guijarro}, C., {Toft}, S., {Karim}, A., {et~al.} 2018, \apj, 856,
  121

\bibitem[{{Gullberg} {et~al.}(2019){Gullberg}, {Smail}, {Swinbank},
  {Dudzevi{\v{c}}i{\={u}}t{\.{e}}}, {Stach}, {Thomson}, {Almaini}, {Chen},
  {Conselice}, {Cooke}, {Farrah}, {Ivison}, {Maltby}, {Micha{\l}owski},
  {Simpson}, {Scott}, {Wardlow}, \& {Weiss}}]{Gullberg2019}
{Gullberg}, B., {Smail}, I., {Swinbank}, A.~M., {et~al.} 2019, \mnras, 490,
  4956

\bibitem[{{Gullberg} {et~al.}(2018){Gullberg}, {Swinbank}, {Smail}, {Biggs},
  {Bertoldi}, {De Breuck}, {Chapman}, {Chen}, {Cooke}, {Coppin}, {Cox},
  {Dannerbauer}, {Dunlop}, {Edge}, {Farrah}, {Geach}, {Greve}, {Hodge}, {Ibar},
  {Ivison}, {Karim}, {Schinnerer}, {Scott}, {Simpson}, {Stach}, {Thomson}, {van
  der Werf}, {Walter}, {Wardlow}, \& {Weiss}}]{Gullberg2018}
{Gullberg}, B., {Swinbank}, A.~M., {Smail}, I., {et~al.} 2018, \apj, 859, 12

\bibitem[{{Harrington} {et~al.}(2021){Harrington}, {Weiss}, {Yun}, {Magnelli},
  {Sharon}, {Leung}, {Vishwas}, {Wang}, {Frayer}, {Jim{\'e}nez-Andrade}, {Liu},
  {Garc{\'\i}a}, {Romano-D{\'\i}az}, {Frye}, {Jarugula}, {B{\u{a}}descu},
  {Berman}, {Dannerbauer}, {D{\'\i}az-S{\'a}nchez}, {Grassitelli},
  {Kamieneski}, {Kim}, {Kirkpatrick}, {Lowenthal}, {Messias}, {Puschnig},
  {Stacey}, {Torne}, \& {Bertoldi}}]{Harrington2021}
{Harrington}, K.~C., {Weiss}, A., {Yun}, M.~S., {et~al.} 2021, \apj, 908, 95

\bibitem[{{H{\"a}u{\ss}ler} {et~al.}(2013){H{\"a}u{\ss}ler}, {Bamford}, {Vika},
  {Rojas}, {Barden}, {Kelvin}, {Alpaslan}, {Robotham}, {Driver}, {Baldry},
  {Brough}, {Hopkins}, {Liske}, {Nichol}, {Popescu}, \& {Tuffs}}]{Haubler2013}
{H{\"a}u{\ss}ler}, B., {Bamford}, S.~P., {Vika}, M., {et~al.} 2013, \mnras,
  430, 330

\bibitem[{{H{\"a}u{\ss}ler} {et~al.}(2022){H{\"a}u{\ss}ler}, {Vika}, {Bamford},
  {Johnston}, {Brough}, {Casura}, {Holwerda}, {Kelvin}, \&
  {Popescu}}]{Haubler2022}
{H{\"a}u{\ss}ler}, B., {Vika}, M., {Bamford}, S.~P., {et~al.} 2022, \aap, 664,
  A92

\bibitem[{{Hayward} {et~al.}(2021){Hayward}, {Sparre}, {Chapman}, {Hernquist},
  {Nelson}, {Pakmor}, {Pillepich}, {Springel}, {Torrey}, {Vogelsberger}, \&
  {Weinberger}}]{Hayward2021}
{Hayward}, C.~C., {Sparre}, M., {Chapman}, S.~C., {et~al.} 2021, \mnras, 502,
  2922

\bibitem[{{Head} {et~al.}(2014){Head}, {Lucey}, {Hudson}, \&
  {Smith}}]{Head2014}
{Head}, J. T.~C.~G., {Lucey}, J.~R., {Hudson}, M.~J., \& {Smith}, R.~J. 2014,
  \mnras, 440, 1690

\bibitem[{{Hill} {et~al.}(2018){Hill}, {Chapman}, {Scott}, {Petitpas}, {Smail},
  {Chapin}, {Gurwell}, {Perry}, {Blain}, {Bremer}, {Chen}, {Dunlop}, {Farrah},
  {Fazio}, {Geach}, {Howson}, {Ivison}, {Lacaille}, {Micha{\l}owski},
  {Simpson}, {Swinbank}, {van der Werf}, \& {Wilner}}]{Hill2018}
{Hill}, R., {Chapman}, S.~C., {Scott}, D., {et~al.} 2018, \mnras, 477, 2042

\bibitem[{{Hodge} {et~al.}(2012){Hodge}, {Carilli}, {Walter}, {de Blok},
  {Riechers}, {Daddi}, \& {Lentati}}]{Hodge2012}
{Hodge}, J.~A., {Carilli}, C.~L., {Walter}, F., {et~al.} 2012, \apj, 760, 11

\bibitem[{{Hodge} \& {da Cunha}(2020)}]{Hodge2020}
{Hodge}, J.~A. \& {da Cunha}, E. 2020, Royal Society Open Science, 7, 200556

\bibitem[{{Hodge} {et~al.}(2019){Hodge}, {Smail}, {Walter}, {da Cunha},
  {Swinbank}, {Rybak}, {Venemans}, {Brandt}, {Calistro Rivera}, {Chapman},
  {Chen}, {Cox}, {Dannerbauer}, {Decarli}, {Greve}, {Knudsen}, {Menten},
  {Schinnerer}, {Simpson}, {van der Werf}, {Wardlow}, \& {Weiss}}]{Hodge2019}
{Hodge}, J.~A., {Smail}, I., {Walter}, F., {et~al.} 2019, \apj, 876, 130

\bibitem[{{Hodge} {et~al.}(2016){Hodge}, {Swinbank}, {Simpson}, {Smail},
  {Walter}, {Alexander}, {Bertoldi}, {Biggs}, {Brandt}, {Chapman}, {Chen},
  {Coppin}, {Cox}, {Dannerbauer}, {Edge}, {Greve}, {Ivison}, {Karim},
  {Knudsen}, {Menten}, {Rix}, {Schinnerer}, {Wardlow}, {Weiss}, \& {van der
  Werf}}]{Hodge2016}
{Hodge}, J.~A., {Swinbank}, A.~M., {Simpson}, J.~M., {et~al.} 2016, \apj, 833,
  103

\bibitem[{{Holland} {et~al.}(2013){Holland}, {Bintley}, {Chapin},
  {Chrysostomou}, {Davis}, {Dempsey}, {Duncan}, {Fich}, {Friberg}, {Halpern},
  {Irwin}, {Jenness}, {Kelly}, {MacIntosh}, {Robson}, {Scott}, {Ade},
  {Atad-Ettedgui}, {Berry}, {Craig}, {Gao}, {Gibb}, {Hilton}, {Hollister},
  {Kycia}, {Lunney}, {McGregor}, {Montgomery}, {Parkes}, {Tilanus}, {Ullom},
  {Walther}, {Walton}, {Woodcraft}, {Amiri}, {Atkinson}, {Burger}, {Chuter},
  {Coulson}, {Doriese}, {Dunare}, {Economou}, {Niemack}, {Parsons},
  {Reintsema}, {Sibthorpe}, {Smail}, {Sudiwala}, \& {Thomas}}]{Holland2013}
{Holland}, W.~S., {Bintley}, D., {Chapin}, E.~L., {et~al.} 2013, \mnras, 430,
  2513

\bibitem[{{Ivison} {et~al.}(2007){Ivison}, {Greve}, {Dunlop}, {Peacock},
  {Egami}, {Smail}, {Ibar}, {van Kampen}, {Aretxaga}, {Babbedge}, {Biggs},
  {Blain}, {Chapman}, {Clements}, {Coppin}, {Farrah}, {Halpern}, {Hughes},
  {Jarvis}, {Jenness}, {Jones}, {Mortier}, {Oliver}, {Papovich},
  {P{\'e}rez-Gonz{\'a}lez}, {Pope}, {Rawlings}, {Rieke}, {Rowan-Robinson},
  {Savage}, {Scott}, {Seigar}, {Serjeant}, {Simpson}, {Stevens}, {Vaccari},
  {Wagg}, \& {Willott}}]{Ivison2007}
{Ivison}, R.~J., {Greve}, T.~R., {Dunlop}, J.~S., {et~al.} 2007, \mnras, 380,
  199

\bibitem[{{Karim} {et~al.}(2013){Karim}, {Swinbank}, {Hodge}, {Smail},
  {Walter}, {Biggs}, {Simpson}, {Danielson}, {Alexander}, {Bertoldi}, {de
  Breuck}, {Chapman}, {Coppin}, {Dannerbauer}, {Edge}, {Greve}, {Ivison},
  {Knudsen}, {Menten}, {Schinnerer}, {Wardlow}, {Wei{\ss}}, \& {van der
  Werf}}]{Karim2013}
{Karim}, A., {Swinbank}, A.~M., {Hodge}, J.~A., {et~al.} 2013, \mnras, 432, 2

\bibitem[{{Kartaltepe} {et~al.}(2022){Kartaltepe}, {Rose}, {Vanderhoof},
  {McGrath}, {Costantin}, {Cox}, {Yung}, {Kocevski}, {Wuyts}, {Andrews},
  {Bagley}, {Finkelstein}, {Amorin}, {Arrabal Haro}, {Backhaus}, {Behroozi},
  {Bisigello}, {Calabro}, {Casey}, {Coogan}, {Croton}, {de la Vega},
  {Dickinson}, {Cooper}, {Fontana}, {Franco}, {Grazian}, {Grogin}, {Hathi},
  {Holwerda}, {Huertas-Company}, {Iyer}, {Jogee}, {Jung}, {Kewley},
  {Kirkpatrick}, {Koekemoer}, {Liu}, {Lotz}, {Lucas}, {Newman}, {Pacifici},
  {Pandya}, {Papovich}, {Pentericci}, {Perez-Gonzalez}, {Petersen}, {Pirzkal},
  {Rafelski}, {Ravindranath}, {Simons}, {Snyder}, {Somerville}, {Stanway},
  {Straughn}, {Tacchella}, {Trump}, {Vega-Ferrero}, {Wilkins}, {Yang}, \&
  {Zavala}}]{Kart2022}
{Kartaltepe}, J.~S., {Rose}, C., {Vanderhoof}, B.~N., {et~al.} 2022, arXiv
  e-prints, arXiv:2210.14713

\bibitem[{{Kokorev} {et~al.}(2022){Kokorev}, {Brammer}, {Fujimoto}, {Kohno},
  {Magdis}, {Valentino}, {Toft}, {Oesch}, {Davidzon}, {Bauer}, {Coe}, {Egami},
  {Oguri}, {Ouchi}, {Postman}, {Richard}, {Jolly}, {Knudsen}, {Sun}, {Weaver},
  {Ao}, {Baker}, {Bradley}, {Caputi}, {Dessauges-Zavadsky}, {Espada},
  {Hatsukade}, {Koekemoer}, {Mu{\~n}oz Arancibia}, {Shimasaku}, {Umehata},
  {Wang}, \& {Wang}}]{Kokorev2022}
{Kokorev}, V., {Brammer}, G., {Fujimoto}, S., {et~al.} 2022, \apjs, 263, 38

\bibitem[{{Krick} {et~al.}(2021){Krick}, {Lowrance}, {Carey}, {Laine},
  {Grillmair}, {Van Dyk}, {Glaccum}, {Ingalls}, {Rieke}, {Hora}, {Fazio},
  {Gordon}, \& {Bohlin}}]{Krick2021}
{Krick}, J.~E., {Lowrance}, P., {Carey}, S., {et~al.} 2021, \aj, 161, 177

\bibitem[{{Kron}(1980)}]{Kron1980}
{Kron}, R.~G. 1980, \apjs, 43, 305

\bibitem[{{Lang} {et~al.}(2019){Lang}, {Schinnerer}, {Smail},
  {Dudzevi{\v{c}}i{\={u}}t{\.{e}}}, {Swinbank}, {Liu}, {Leslie}, {Almaini},
  {An}, {Bertoldi}, {Blain}, {Chapman}, {Chen}, {Conselice}, {Cooke}, {Coppin},
  {Dunlop}, {Farrah}, {Fudamoto}, {Geach}, {Gullberg}, {Harrington}, {Hodge},
  {Ivison}, {Jim{\'e}nez-Andrade}, {Magnelli}, {Micha{\l}owski}, {Oesch},
  {Scott}, {Simpson}, {Smol{\v{c}}i{\'c}}, {Stach}, {Thomson}, {Toft},
  {Vardoulaki}, {Wardlow}, {Weiss}, \& {van der Werf}}]{Lang2019}
{Lang}, P., {Schinnerer}, E., {Smail}, I., {et~al.} 2019, \apj, 879, 54

\bibitem[{{Lange} {et~al.}(2016){Lange}, {Moffett}, {Driver}, {Robotham},
  {Lagos}, {Kelvin}, {Conselice}, {Margalef-Bentabol}, {Alpaslan}, {Baldry},
  {Bland-Hawthorn}, {Bremer}, {Brough}, {Cluver}, {Colless}, {Davies},
  {H{\"a}u{\ss}ler}, {Holwerda}, {Hopkins}, {Kafle}, {Kennedy}, {Liske},
  {Phillipps}, {Popescu}, {Taylor}, {Tuffs}, {van Kampen}, \&
  {Wright}}]{Lange2016}
{Lange}, R., {Moffett}, A.~J., {Driver}, S.~P., {et~al.} 2016, \mnras, 462,
  1470

\bibitem[{{Law} {et~al.}(2012){Law}, {Steidel}, {Shapley}, {Nagy}, {Reddy}, \&
  {Erb}}]{Law2012}
{Law}, D.~R., {Steidel}, C.~C., {Shapley}, A.~E., {et~al.} 2012, \apj, 745, 85

\bibitem[{{Lelli} {et~al.}(2021){Lelli}, {Di Teodoro}, {Fraternali}, {Man},
  {Zhang}, {De Breuck}, {Davis}, \& {Maiolino}}]{Lelli2021}
{Lelli}, F., {Di Teodoro}, E.~M., {Fraternali}, F., {et~al.} 2021, Science,
  371, 713

\bibitem[{{Liddle}(2007)}]{Liddle2007}
{Liddle}, A.~R. 2007, \mnras, 377, L74

\bibitem[{{Ling} \& {Yan}(2022)}]{Ling2022}
{Ling}, C. \& {Yan}, H. 2022, \apj, 929, 40

\bibitem[{{Long} {et~al.}(2022){Long}, {Casey}, {Lagos}, {Lambrides}, {Zavala},
  {Champagne}, {Cooper}, \& {Cooray}}]{Long2022}
{Long}, A.~S., {Casey}, C.~M., {Lagos}, C. d.~P., {et~al.} 2022, arXiv
  e-prints, arXiv:2211.02072

\bibitem[{{Lotz} {et~al.}(2008){Lotz}, {Davis}, {Faber}, {Guhathakurta},
  {Gwyn}, {Huang}, {Koo}, {Le Floc'h}, {Lin}, {Newman}, {Noeske}, {Papovich},
  {Willmer}, {Coil}, {Conselice}, {Cooper}, {Hopkins}, {Metevier}, {Primack},
  {Rieke}, \& {Weiner}}]{Lotz2008}
{Lotz}, J.~M., {Davis}, M., {Faber}, S.~M., {et~al.} 2008, \apj, 672, 177

\bibitem[{{Lotz} {et~al.}(2004){Lotz}, {Primack}, \& {Madau}}]{Lotz2004}
{Lotz}, J.~M., {Primack}, J., \& {Madau}, P. 2004, \aj, 128, 163

\bibitem[{{Lovell} {et~al.}(2021){Lovell}, {Geach}, {Dav{\'e}}, {Narayanan}, \&
  {Li}}]{Lovell2021}
{Lovell}, C.~C., {Geach}, J.~E., {Dav{\'e}}, R., {Narayanan}, D., \& {Li}, Q.
  2021, \mnras, 502, 772

\bibitem[{{Lower} {et~al.}(2022){Lower}, {Narayanan}, {Li}, \&
  {Dav{\'e}}}]{Lower2022}
{Lower}, S., {Narayanan}, D., {Li}, Q., \& {Dav{\'e}}, R. 2022, arXiv e-prints,
  arXiv:2212.02636

\bibitem[{{Manning} {et~al.}(2022){Manning}, {Casey}, {Zavala}, {Magdis},
  {Drew}, {Champagne}, {Aravena}, {B{\'e}thermin}, {Clements}, {Finkelstein},
  {Fujimoto}, {Hayward}, {Hodge}, {Ilbert}, {Kartaltepe}, {Knudsen},
  {Koekemoer}, {Man}, {Sanders}, {Sheth}, {Spilker}, {Staguhn}, {Talia},
  {Treister}, \& {Yun}}]{Manning2022}
{Manning}, S.~M., {Casey}, C.~M., {Zavala}, J.~A., {et~al.} 2022, \apj, 925, 23

\bibitem[{{Martis} {et~al.}(2022){Martis}, {Marchesini}, {Muzzin}, {Willott},
  \& {Sawicki}}]{Martis2022}
{Martis}, N.~S., {Marchesini}, D.~M., {Muzzin}, A., {Willott}, C.~J., \&
  {Sawicki}, M. 2022, \mnras [\eprint[arXiv]{2211.12662}]

\bibitem[{{McAlpine} {et~al.}(2019){McAlpine}, {Smail}, {Bower}, {Swinbank},
  {Trayford}, {Theuns}, {Baes}, {Camps}, {Crain}, \& {Schaye}}]{Mcalpine2019}
{McAlpine}, S., {Smail}, I., {Bower}, R.~G., {et~al.} 2019, \mnras, 488, 2440

\bibitem[{{Miller} {et~al.}(2018){Miller}, {Chapman}, {Aravena}, {Ashby},
  {Hayward}, {Vieira}, {Wei{\ss}}, {Babul}, {B{\'e}thermin}, {Bradford},
  {Brodwin}, {Carlstrom}, {Chen}, {Cunningham}, {De Breuck}, {Gonzalez},
  {Greve}, {Harnett}, {Hezaveh}, {Lacaille}, {Litke}, {Ma}, {Malkan},
  {Marrone}, {Morningstar}, {Murphy}, {Narayanan}, {Pass}, {Perry}, {Phadke},
  {Rennehan}, {Rotermund}, {Simpson}, {Spilker}, {Sreevani}, {Stark},
  {Strandet}, \& {Strom}}]{Miller2018}
{Miller}, T.~B., {Chapman}, S.~C., {Aravena}, M., {et~al.} 2018, \nat, 556, 469

\bibitem[{{Nardiello} {et~al.}(2022){Nardiello}, {Bedin}, {Burgasser},
  {Salaris}, {Cassisi}, {Griggio}, \& {Scalco}}]{Nardiello2022}
{Nardiello}, D., {Bedin}, L.~R., {Burgasser}, A., {et~al.} 2022, \mnras, 517,
  484

\bibitem[{{Nelson} {et~al.}(2019){Nelson}, {Pillepich}, {Springel}, {Pakmor},
  {Weinberger}, {Genel}, {Torrey}, {Vogelsberger}, {Marinacci}, \&
  {Hernquist}}]{Nelson2019}
{Nelson}, D., {Pillepich}, A., {Springel}, V., {et~al.} 2019, \mnras, 490, 3234

\bibitem[{{Nelson} {et~al.}(2022){Nelson}, {Suess}, {Bezanson}, {Price}, {van
  Dokkum}, {Leja}, {Wang}, {Whitaker}, {Labb{\'e}}, {Barrufet}, {Brammer},
  {Eisenstein}, {Heintz}, {Johnson}, {Mathews}, {Miller}, {Oesch}, {Sandles},
  {Setton}, {Speagle}, {Tacchella}, {Tadaki}, \& {Weaver}}]{Nelson2022}
{Nelson}, E.~J., {Suess}, K.~A., {Bezanson}, R., {et~al.} 2022, arXiv e-prints,
  arXiv:2208.01630

\bibitem[{{Oteo} {et~al.}(2018){Oteo}, {Ivison}, {Dunne}, {Manilla-Robles},
  {Maddox}, {Lewis}, {de Zotti}, {Bremer}, {Clements}, {Cooray}, {Dannerbauer},
  {Eales}, {Greenslade}, {Omont}, {Perez{\textendash}Fourn{\'o}n}, {Riechers},
  {Scott}, {van der Werf}, {Weiss}, \& {Zhang}}]{Oteo2018}
{Oteo}, I., {Ivison}, R.~J., {Dunne}, L., {et~al.} 2018, \apj, 856, 72

\bibitem[{{Pacifici} {et~al.}(2022){Pacifici}, {Iyer}, {Mobasher}, {da Cunha},
  {Acquaviva}, {Burgarella}, {Calistro Rivera}, {Carnall}, {Chang}, {Chartab},
  {Cooke}, {Fairhurst}, {Kartaltepe}, {Leja}, {Malek}, {Salmon}, {Torelli},
  {Vidal-Garcia}, {Boquien}, {Brammer}, {Brown}, {Capak}, {Chevallard},
  {Circosta}, {Croton}, {Davidzon}, {Dickinson}, {Duncan}, {Faber}, {Ferguson},
  {Fontana}, {Guo}, {Haeussler}, {Hemmati}, {Jafariyazani}, {Kassin}, {Larson},
  {Lee}, {Mantha}, {Marchi}, {Nayyeri}, {Newman}, {Pandya}, {Pforr}, {Reddy},
  {Sanders}, {Shah}, {Shahidi}, {Stevans}, {Puspita Triani}, {Tyler},
  {Vanderhoof}, {de la Vega}, {Wang}, \& {Weston}}]{Pac2022}
{Pacifici}, C., {Iyer}, K.~G., {Mobasher}, B., {et~al.} 2022, arXiv e-prints,
  arXiv:2212.01915

\bibitem[{{Peng} {et~al.}(2010){Peng}, {Ho}, {Impey}, \& {Rix}}]{Peng2010}
{Peng}, C.~Y., {Ho}, L.~C., {Impey}, C.~D., \& {Rix}, H.-W. 2010, \aj, 139,
  2097

\bibitem[{{Perrin} {et~al.}(2014){Perrin}, {Sivaramakrishnan}, {Lajoie},
  {Elliott}, {Pueyo}, {Ravindranath}, \& {Albert}}]{Perrin2014}
{Perrin}, M.~D., {Sivaramakrishnan}, A., {Lajoie}, C.-P., {et~al.} 2014, in
  Society of Photo-Optical Instrumentation Engineers (SPIE) Conference Series,
  Vol. 9143, Space Telescopes and Instrumentation 2014: Optical, Infrared, and
  Millimeter Wave, ed. J.~{Oschmann}, Jacobus~M., M.~{Clampin}, G.~G. {Fazio},
  \& H.~A. {MacEwen}, 91433X

\bibitem[{{Popping} {et~al.}(2022){Popping}, {Pillepich}, {Calistro Rivera},
  {Schulz}, {Hernquist}, {Kaasinen}, {Marinacci}, {Nelson}, \&
  {Vogelsberger}}]{Popping2022}
{Popping}, G., {Pillepich}, A., {Calistro Rivera}, G., {et~al.} 2022, \mnras,
  510, 3321

\bibitem[{{Rizzo} {et~al.}(2021){Rizzo}, {Vegetti}, {Fraternali}, {Stacey}, \&
  {Powell}}]{Rizzo2021}
{Rizzo}, F., {Vegetti}, S., {Fraternali}, F., {Stacey}, H.~R., \& {Powell}, D.
  2021, \mnras, 507, 3952

\bibitem[{{Robertson} {et~al.}(2022){Robertson}, {Tacchella}, {Johnson},
  {Hausen}, {Alabi}, {Boyett}, {Bunker}, {Carniani}, {Egami}, {Eisenstein},
  {Hainline}, {Helton}, {Ji}, {Kumari}, {Lyu}, {Maiolino}, {Nelson}, {Rieke},
  {Shivaei}, {Sun}, {Ubler}, {Williams}, {Willmer}, \&
  {Witstok}}]{Robertson2022}
{Robertson}, B.~E., {Tacchella}, S., {Johnson}, B.~D., {et~al.} 2022, arXiv
  e-prints, arXiv:2208.11456

\bibitem[{{Rodriguez-Gomez} {et~al.}(2019){Rodriguez-Gomez}, {Snyder}, {Lotz},
  {Nelson}, {Pillepich}, {Springel}, {Genel}, {Weinberger}, {Tacchella},
  {Pakmor}, {Torrey}, {Marinacci}, {Vogelsberger}, {Hernquist}, \&
  {Thilker}}]{statmorph2019}
{Rodriguez-Gomez}, V., {Snyder}, G.~F., {Lotz}, J.~M., {et~al.} 2019, \mnras,
  483, 4140

\bibitem[{{Roper} {et~al.}(2023){Roper}, {Lovell}, {Vijayan}, {Irodotou},
  {Kuusisto}, {Matharu}, {Seeyave}, {Thomas}, \& {Wilkins}}]{Roper2023}
{Roper}, W.~J., {Lovell}, C.~C., {Vijayan}, A.~P., {et~al.} 2023, arXiv
  e-prints, arXiv:2301.05228

\bibitem[{{Roper} {et~al.}(2022){Roper}, {Lovell}, {Vijayan}, {Marshall},
  {Irodotou}, {Kuusisto}, {Thomas}, \& {Wilkins}}]{Roper2022}
{Roper}, W.~J., {Lovell}, C.~C., {Vijayan}, A.~P., {et~al.} 2022, \mnras, 514,
  1921

\bibitem[{{Shim} {et~al.}(2022){Shim}, {Lee}, {Kim}, {Scott}, {Serjeant}, {Ao},
  {Barrufet}, {Chapman}, {Clements}, {Conselice}, {Goto}, {Greve}, {Hwang},
  {Im}, {Jeong}, {Kim}, {Kim}, {Kim}, {Kong}, {Koprowski}, {Malkan},
  {Micha{\l}owski}, {Pearson}, {Seo}, {Takagi}, {Toba}, {White}, \&
  {Woo}}]{Shim2022}
{Shim}, H., {Lee}, D., {Kim}, Y., {et~al.} 2022, \mnras, 514, 2915

\bibitem[{{Simpson} {et~al.}(2014){Simpson}, {Swinbank}, {Smail}, {Alexander},
  {Brandt}, {Bertoldi}, {de Breuck}, {Chapman}, {Coppin}, {da Cunha},
  {Danielson}, {Dannerbauer}, {Greve}, {Hodge}, {Ivison}, {Karim}, {Knudsen},
  {Poggianti}, {Schinnerer}, {Thomson}, {Walter}, {Wardlow}, {Wei{\ss}}, \&
  {van der Werf}}]{Simpson2014}
{Simpson}, J.~M., {Swinbank}, A.~M., {Smail}, I., {et~al.} 2014, \apj, 788, 125

\bibitem[{{Smail} {et~al.}(2021){Smail}, {Dudzevi{\v{c}}i{\={u}}t{\.{e}}},
  {Stach}, {Almaini}, {Birkin}, {Chapman}, {Chen}, {Geach}, {Gullberg},
  {Hodge}, {Ikarashi}, {Ivison}, {Scott}, {Simpson}, {Swinbank}, {Thomson},
  {Walter}, {Wardlow}, \& {van der Werf}}]{Smail2021}
{Smail}, I., {Dudzevi{\v{c}}i{\={u}}t{\.{e}}}, U., {Stach}, S.~M., {et~al.}
  2021, \mnras, 502, 3426

\bibitem[{{Smail} {et~al.}(1997){Smail}, {Ivison}, \& {Blain}}]{Smail1997}
{Smail}, I., {Ivison}, R.~J., \& {Blain}, A.~W. 1997, \apjl, 490, L5

\bibitem[{{Snyder} {et~al.}(2015){Snyder}, {Torrey}, {Lotz}, {Genel},
  {McBride}, {Vogelsberger}, {Pillepich}, {Nelson}, {Sales}, {Sijacki},
  {Hernquist}, \& {Springel}}]{Synder2015}
{Snyder}, G.~F., {Torrey}, P., {Lotz}, J.~M., {et~al.} 2015, \mnras, 454, 1886

\bibitem[{{Stach} {et~al.}(2019){Stach}, {Dudzevi{\v{c}}i{\={u}}t{\.{e}}},
  {Smail}, {Swinbank}, {Geach}, {Simpson}, {An}, {Almaini}, {Arumugam},
  {Blain}, {Chapman}, {Chen}, {Conselice}, {Cooke}, {Coppin}, {da Cunha},
  {Dunlop}, {Farrah}, {Gullberg}, {Hodge}, {Ivison}, {Kocevski},
  {Micha{\l}owski}, {Miyaji}, {Scott}, {Thomson}, {Wardlow}, {Weiss}, \& {van
  der Werf}}]{Stach2019}
{Stach}, S.~M., {Dudzevi{\v{c}}i{\={u}}t{\.{e}}}, U., {Smail}, I., {et~al.}
  2019, \mnras, 487, 4648

\bibitem[{{Stach} {et~al.}(2018){Stach}, {Smail}, {Swinbank}, {Simpson},
  {Geach}, {An}, {Almaini}, {Arumugam}, {Blain}, {Chapman}, {Chen},
  {Conselice}, {Cooke}, {Coppin}, {Dunlop}, {Farrah}, {Gullberg}, {Hartley},
  {Ivison}, {Maltby}, {Micha{\l}owski}, {Scott}, {Simpson}, {Thomson},
  {Wardlow}, \& {van der Werf}}]{Stach2018}
{Stach}, S.~M., {Smail}, I., {Swinbank}, A.~M., {et~al.} 2018, \apj, 860, 161

\bibitem[{{Swinbank} {et~al.}(2014){Swinbank}, {Simpson}, {Smail}, {Harrison},
  {Hodge}, {Karim}, {Walter}, {Alexander}, {Brandt}, {de Breuck}, {da Cunha},
  {Chapman}, {Coppin}, {Danielson}, {Dannerbauer}, {Decarli}, {Greve},
  {Ivison}, {Knudsen}, {Lagos}, {Schinnerer}, {Thomson}, {Wardlow}, {Wei{\ss}},
  \& {van der Werf}}]{Swinbank2014}
{Swinbank}, A.~M., {Simpson}, J.~M., {Smail}, I., {et~al.} 2014, \mnras, 438,
  1267

\bibitem[{{Swinbank} {et~al.}(2010){Swinbank}, {Smail}, {Chapman}, {Borys},
  {Alexander}, {Blain}, {Conselice}, {Hainline}, \& {Ivison}}]{Swinbank2010}
{Swinbank}, A.~M., {Smail}, I., {Chapman}, S.~C., {et~al.} 2010, \mnras, 405,
  234

\bibitem[{{van der Wel} {et~al.}(2014{\natexlab{a}}){van der Wel}, {Chang},
  {Bell}, {Holden}, {Ferguson}, {Giavalisco}, {Rix}, {Skelton}, {Whitaker},
  {Momcheva}, {Brammer}, {Kassin}, {Martig}, {Dekel}, {Ceverino}, {Koo},
  {Mozena}, {van Dokkum}, {Franx}, {Faber}, \& {Primack}}]{VW2014}
{van der Wel}, A., {Chang}, Y.-Y., {Bell}, E.~F., {et~al.} 2014{\natexlab{a}},
  \apjl, 792, L6

\bibitem[{{van der Wel} {et~al.}(2014{\natexlab{b}}){van der Wel}, {Franx},
  {van Dokkum}, {Skelton}, {Momcheva}, {Whitaker}, {Brammer}, {Bell}, {Rix},
  {Wuyts}, {Ferguson}, {Holden}, {Barro}, {Koekemoer}, {Chang}, {McGrath},
  {H{\"a}ussler}, {Dekel}, {Behroozi}, {Fumagalli}, {Leja}, {Lundgren},
  {Maseda}, {Nelson}, {Wake}, {Patel}, {Labb{\'e}}, {Faber}, {Grogin}, \&
  {Kocevski}}]{VW2014a}
{van der Wel}, A., {Franx}, M., {van Dokkum}, P.~G., {et~al.}
  2014{\natexlab{b}}, \apj, 788, 28

\bibitem[{{Walter} {et~al.}(2012){Walter}, {Decarli}, {Carilli}, {Bertoldi},
  {Cox}, {da Cunha}, {Daddi}, {Dickinson}, {Downes}, {Elbaz}, {Ellis}, {Hodge},
  {Neri}, {Riechers}, {Weiss}, {Bell}, {Dannerbauer}, {Krips}, {Krumholz},
  {Lentati}, {Maiolino}, {Menten}, {Rix}, {Robertson}, {Spinrad}, {Stark}, \&
  {Stern}}]{Walter2012}
{Walter}, F., {Decarli}, R., {Carilli}, C., {et~al.} 2012, \nat, 486, 233

\bibitem[{{Wardlow} {et~al.}(2011){Wardlow}, {Smail}, {Coppin}, {Alexander},
  {Brandt}, {Danielson}, {Luo}, {Swinbank}, {Walter}, {Wei{\ss}}, {Xue},
  {Zibetti}, {Bertoldi}, {Biggs}, {Chapman}, {Dannerbauer}, {Dunlop},
  {Gawiser}, {Ivison}, {Knudsen}, {Kov{\'a}cs}, {Lacey}, {Menten}, {Padilla},
  {Rix}, \& {van der Werf}}]{Wardlow2011}
{Wardlow}, J.~L., {Smail}, I., {Coppin}, K.~E.~K., {et~al.} 2011, \mnras, 415,
  1479

\bibitem[{{Wei{\ss}} {et~al.}(2013){Wei{\ss}}, {De Breuck}, {Marrone},
  {Vieira}, {Aguirre}, {Aird}, {Aravena}, {Ashby}, {Bayliss}, {Benson},
  {B{\'e}thermin}, {Biggs}, {Bleem}, {Bock}, {Bothwell}, {Bradford}, {Brodwin},
  {Carlstrom}, {Chang}, {Chapman}, {Crawford}, {Crites}, {de Haan}, {Dobbs},
  {Downes}, {Fassnacht}, {George}, {Gladders}, {Gonzalez}, {Greve},
  {Halverson}, {Hezaveh}, {High}, {Holder}, {Holzapfel}, {Hoover}, {Hrubes},
  {Husband}, {Keisler}, {Lee}, {Leitch}, {Lueker}, {Luong-Van}, {Malkan},
  {McIntyre}, {McMahon}, {Mehl}, {Menten}, {Meyer}, {Murphy}, {Padin},
  {Plagge}, {Reichardt}, {Rest}, {Rosenman}, {Ruel}, {Ruhl}, {Schaffer},
  {Shirokoff}, {Spilker}, {Stalder}, {Staniszewski}, {Stark}, {Story},
  {Vanderlinde}, {Welikala}, \& {Williamson}}]{Weiss2013}
{Wei{\ss}}, A., {De Breuck}, C., {Marrone}, D.~P., {et~al.} 2013, \apj, 767, 88

\bibitem[{{Wei{\ss}} {et~al.}(2009){Wei{\ss}}, {Ivison}, {Downes}, {Walter},
  {Cirasuolo}, \& {Menten}}]{Weiss2009}
{Wei{\ss}}, A., {Ivison}, R.~J., {Downes}, D., {et~al.} 2009, \apjl, 705, L45

\bibitem[{{Zavala} {et~al.}(2018){Zavala}, {Aretxaga}, {Dunlop},
  {Micha{\l}owski}, {Hughes}, {Bourne}, {Chapin}, {Cowley}, {Farrah}, {Lacey},
  {Targett}, \& {van der Werf}}]{Zavala2018}
{Zavala}, J.~A., {Aretxaga}, I., {Dunlop}, J.~S., {et~al.} 2018, \mnras, 475,
  5585

\bibitem[{{Zavala} {et~al.}(2017){Zavala}, {Aretxaga}, {Geach}, {Hughes},
  {Birkinshaw}, {Chapin}, {Chapman}, {Chen}, {Clements}, {Dunlop}, {Farrah},
  {Ivison}, {Jenness}, {Micha{\l}owski}, {Robson}, {Scott}, {Simpson},
  {Spaans}, \& {van der Werf}}]{Zavala2017}
{Zavala}, J.~A., {Aretxaga}, I., {Geach}, J.~E., {et~al.} 2017, \mnras, 464,
  3369

\bibitem[{{Zhang} {et~al.}(2019){Zhang}, {Primack}, {Faber}, {Koo}, {Dekel},
  {Chen}, {Ceverino}, {Chang}, {Fang}, {Guo}, {Lin}, \& {Wel}}]{Zhang2019}
{Zhang}, H., {Primack}, J.~R., {Faber}, S.~M., {et~al.} 2019, \mnras, 484, 5170

\end{thebibliography}
\bibliographystyle{aa}

\begin{figure*}
    \appendix
    \section{Decision Tree} \label{App:A}
    \centering
    \includegraphics[width=\linewidth]{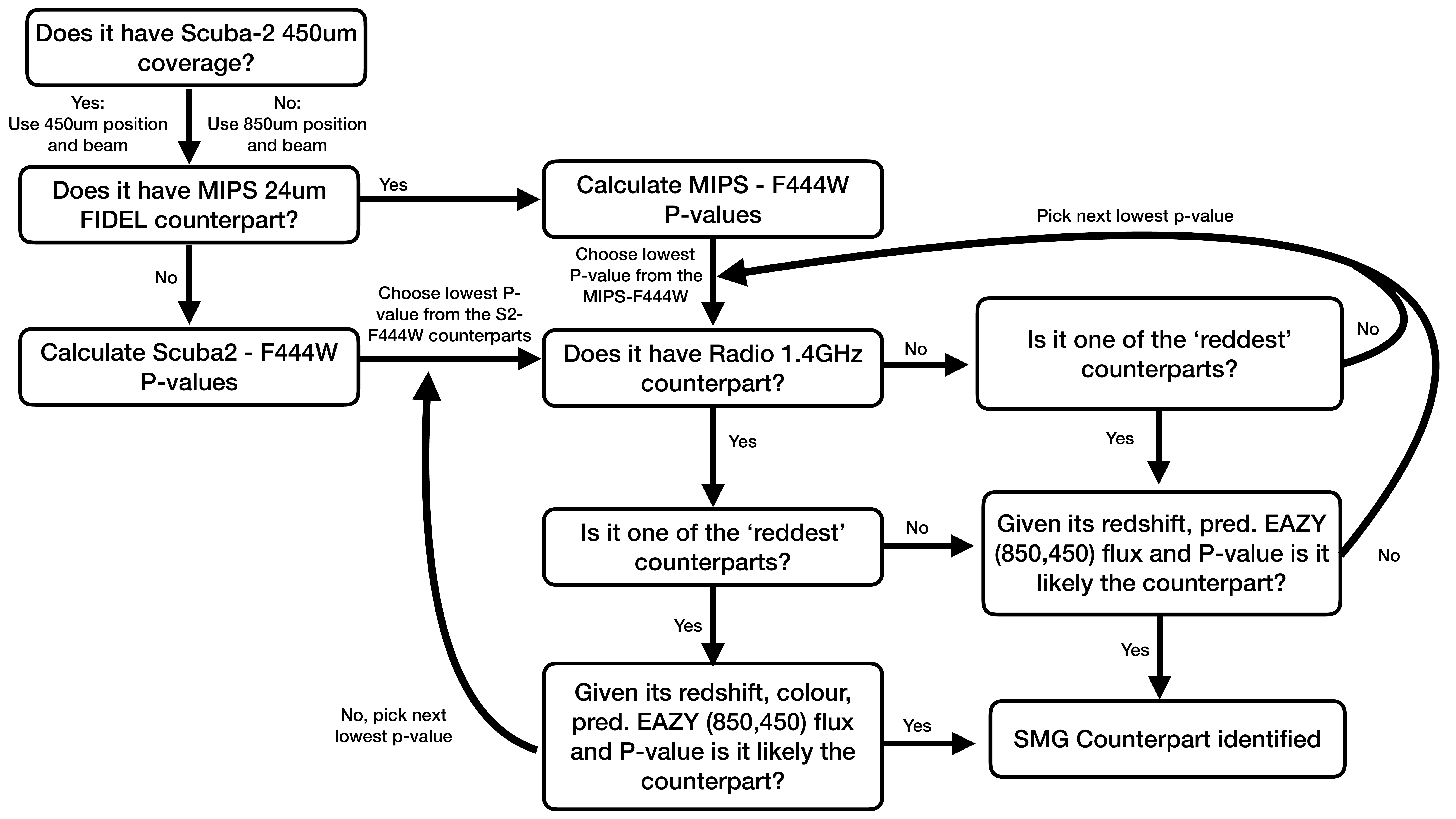}
    \caption{A schematic of the decision tree used to determine the NIRCam counterpart(s) to the SCUBA-2 selected SMGs. If a MIPS counterpart has been identified in the SCUBA-2 beam then $p$-values from the MIPS 24\um\ image to NIRCam F444W are calculated, else directly from SCUBA-2 to NIRCam F444W. Then considering the NIRCam colours and \texttt{EaZy-py} outputs as shown in Figure \ref{Fig:counterparts}, a decision is made.}
     \label{Fig:Tree}
\end{figure*}

\begin{figure*}
\section{\texttt{GALFITM} Modelling}\label{App:galfit}
    \centering
    \includegraphics[width=\linewidth]{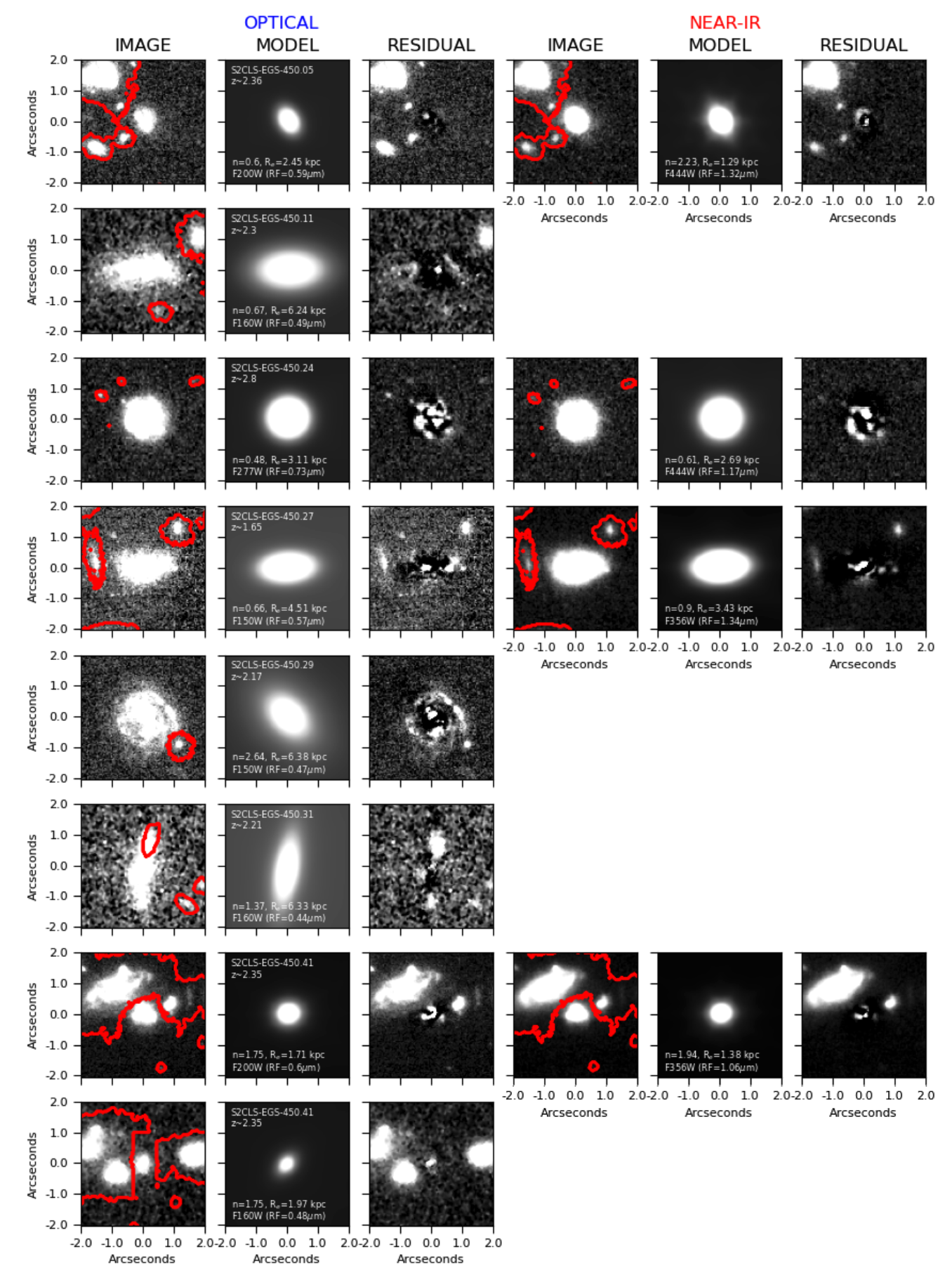}
    \end{figure*}
   \begin{figure*}
    \includegraphics[width=\linewidth]{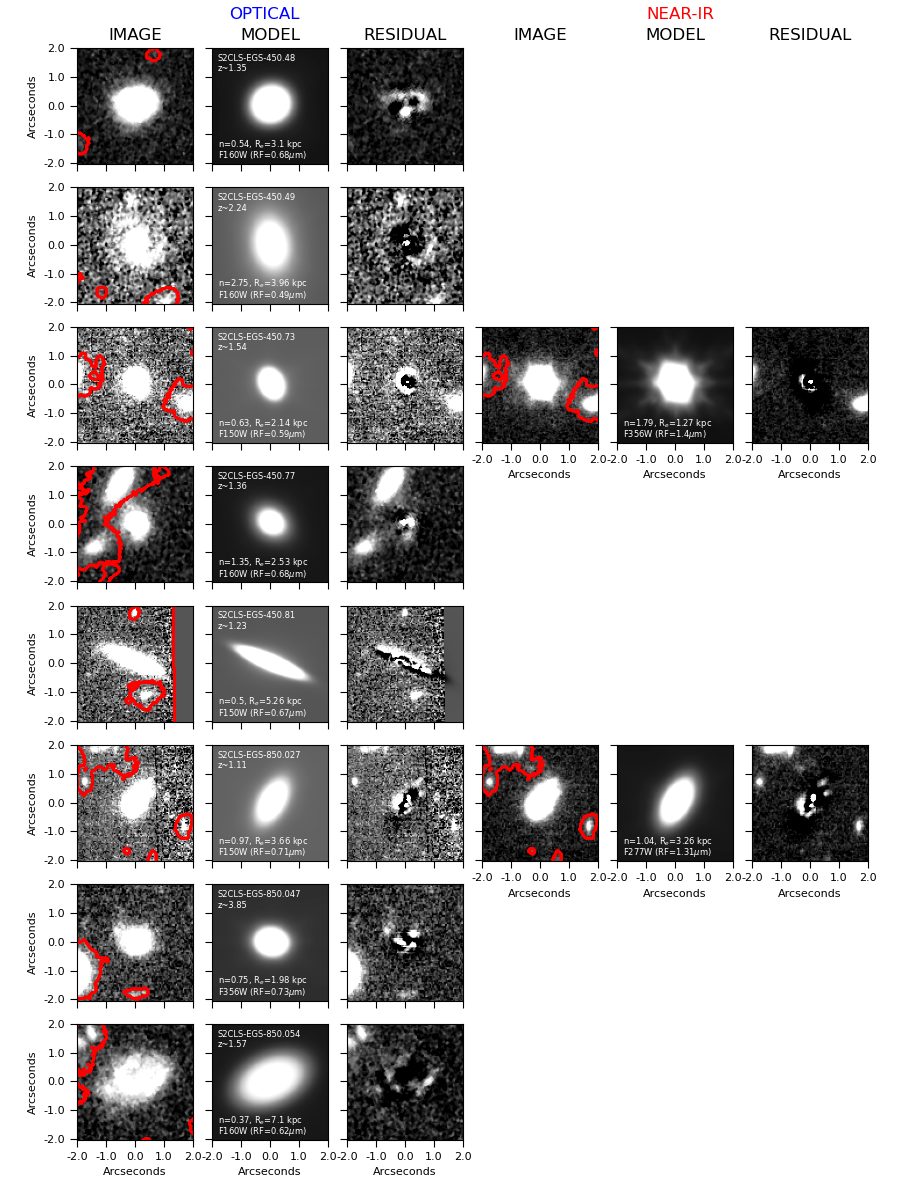}
    \end{figure*}
     \begin{figure*}
    \includegraphics[width=\linewidth]{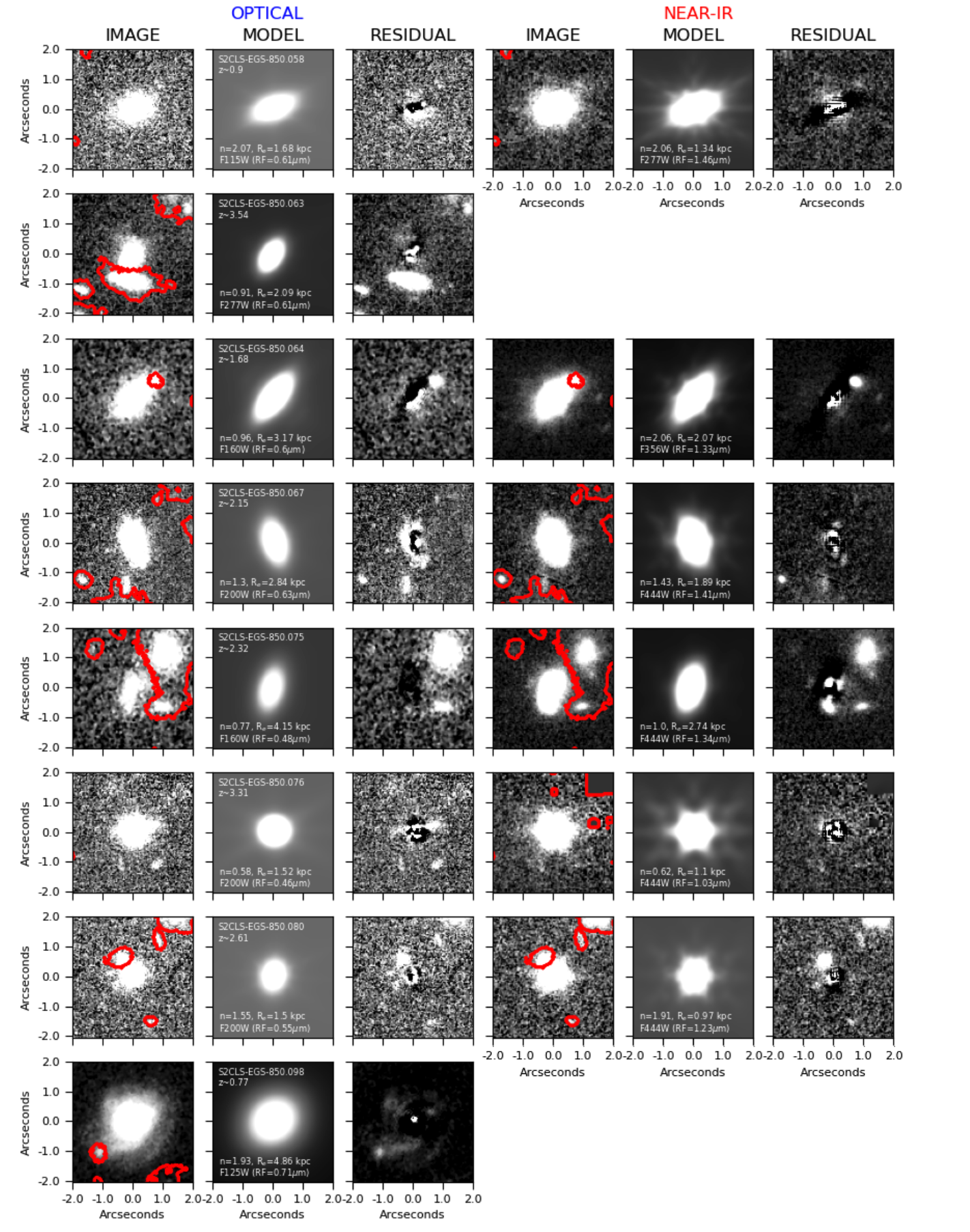}
    \end{figure*}
    \begin{figure*}
    \includegraphics[width=\linewidth,trim={0cm 26cm 0cm 0cm}]{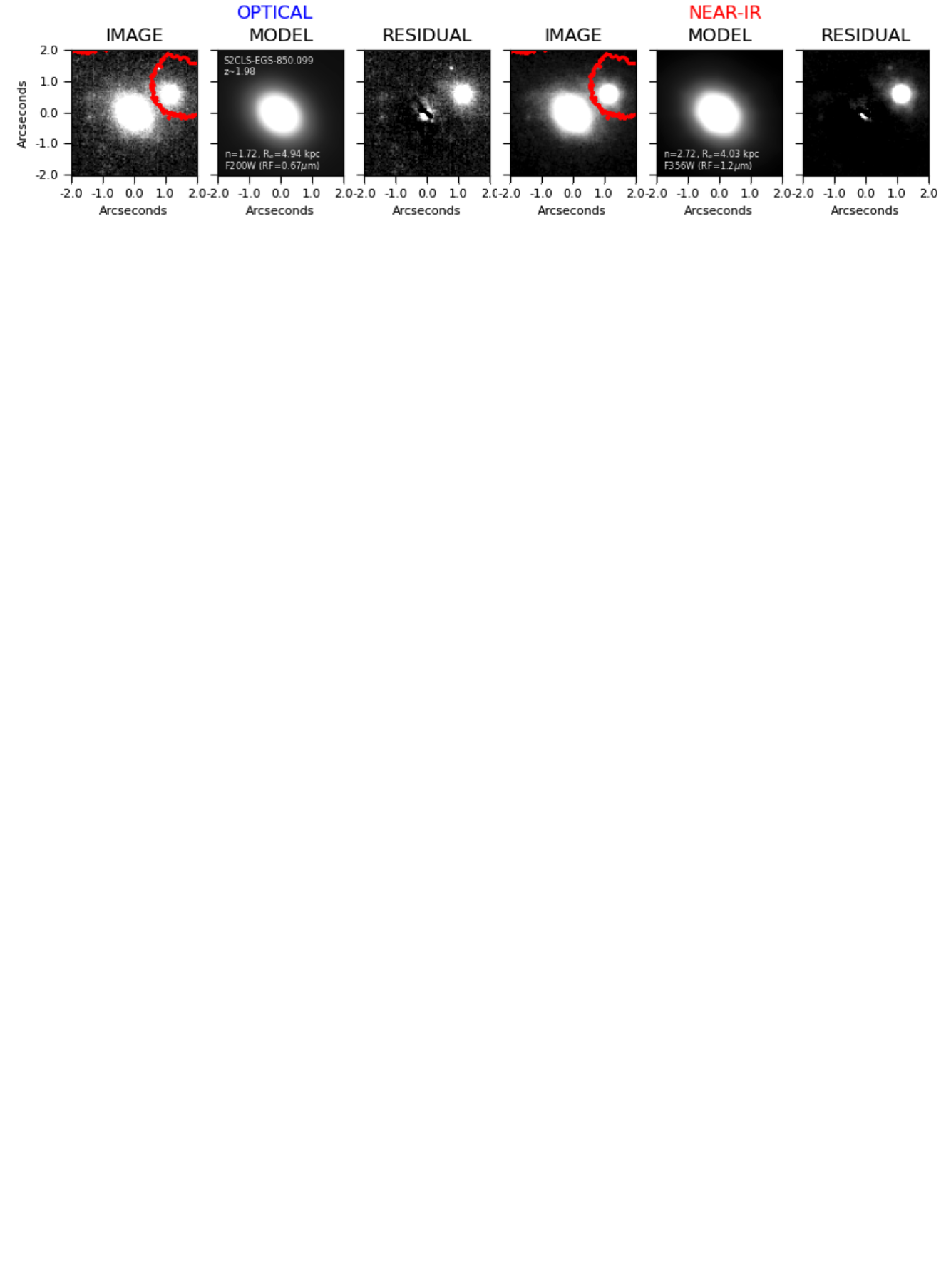}
    \caption{The rest-frame optical and near-infrared \texttt{GALFITM} modelling of the final rest-frame sample presented in Figure~\ref{Fig:rw}. For each NIRCam-SMG we present the 4-arcsecond optical image,  \texttt{GALFITM} model and residual. We further indicate the mask regions of nearby objects (red contours) and indicate for each model the redshift, derived S\'ersic index, effective radius, instrument band and rest-frame wavelength probed. Where available (14/25) we display the near-infrared image, model and residual.}
     \label{Fig:rf_galfitm}
\end{figure*}

\end{document}